# Corotation-bounce resonance of ions in Jupiter's magnetosphere

Y. Sarkango[1], J. R. Szalay[1], P. A. Damiano[2], A. H. Sulaiman[3], P. A. Delamere[2], J. Saur[4], D. J. McComas[1], R. W. Ebert[5,6], F. Allegrini[5,6]

[1]Department of Astrophysical Sciences, Princeton University, USA

[2]Geophysical Institute, University of Alaska Fairbanks, USA

[3]School of Physics and Astronomy, University of Minnesota, USA

[4]Institut für Geophysik und Meteorologie, Universität zu Köln, Germany

[5]Southwest Research Institute, USA

[6]Department of Physics and Astronomy, University of Texas at San Antonio, USA

Corresponding author: Yash Sarkango (sarkango@princeton.edu)

**Key Points:**
- Banded energy distributions (0.01-45 keV/q) occur simultaneously for different ion species mapping to M=10-20.
- Bounce frequencies of banded ions match harmonics of the corotation/System-III frequency.
- Corotation-modulated bounce resonance accelerates low-energy plasma ions in Jupiter's magnetosphere.

**Abstract**

Banded energy distributions of $H^+$, $O^{++}$, $S^{+++}$, and $O^+$ or $S^{++}$ ions between 100 eV to ~20 keV are consistently observed in Jupiter's magnetosphere mapping to M-shells between M=10-20. The bands correspond to flux enhancements at similar speeds for different ion species, providing the first evidence of simultaneous bounce-resonant acceleration of multiple ion species in Jupiter's magnetosphere. Ion enhancements occur for energies at which the bounce frequencies of the trapped ions matched integer harmonics of the System-III corotation frequency. The observations highlight a previously unknown interaction between corotation and bounce motion of <10 keV energy ions that is a fundamental and persistent process occurring in Jupiter's magnetosphere.

**Plain Language Summary**

Trapped plasma particles in Jupiter's magnetosphere experience a "bounce" motion - traveling from the northern to the southern hemisphere and back along magnetic field lines. In addition, plasma as a whole is driven to rotate with the planet, a process referred to as corotation. These are two distinct periodic processes – the former dictates how individual particles travel along a field line, whereas the latter is a global scale process However, for low energy plasma particles, both processes have similar timescales (>100 min). In this work, we show evidence from plasma measurements of resonance between these two processes, which could also be important for plasma acceleration.

**1 Introduction**

Particle flux enhancements at discrete energies, which we refer to as "banded energy distributions," have been observed in ion populations throughout the solar system and have been explained by different processes for the energetic and the thermal populations. They can be classified into at least two categories based on the energy at which bands are observed -

### 1.1. "Zebra-stripes" for energetic particles (> 200 keV electrons and ions)

Energy distributions that appear as zebra-stripes (bands or alternating enhancement and depletions) have been observed for energetic electrons between 40-400 keV and energetic ions at >0.4-1 MeV energies in the magnetospheres of Earth (Lejosne et al., 2022; Lejosne & Mozer, 2020; Pandya et al., 2023, 2024; Sauvaud et al., 2013; Ukhorskiy et al., 2014; Z. Wang et al., 2024), Saturn (Hao et al., 2020; Sun et al., 2021, 2022), and have been predicted for Jupiter (Hao et al., 2020). The physical origins of the zebra-stripe distributions are still under debate. Simulations have reproduced "zebra-stripes" either by using an explicit time-varying electric field (Ukhorskiy et al., 2014), by the influence of a convection electric field (Pandya et al., 2023), by a noon-midnight electric field in the case of Saturn (Sun et al., 2021), or by a dawn-dusk electric field at Jupiter (Hao et al., 2020), without requiring any explicit time-variability in the non-rotating frame. Both at Jupiter and Saturn, a corotation-drift resonance (CDR) mechanism was also proposed in which MeV-energy electrons are quasi-stationary in local time due to gradient-curvature drifts opposing corotation, making them susceptible to other large-scale electric fields (Roussos et al., 2018). Zebra-stripe distributions for energetic particles are not the focus of the present work, we mention them only to avoid confusion.

### 1.2. Banded energy distributions for low-energy particles (<10 keV electrons and ions)

Banded energy distributions have also been reported for lower energy plasma ions, for which the gradient-curvature drift is negligibly slow, and these are likely produced by a different process. For example, the gradient-curvature drift speed for a 10 keV particle (either ion or electron) at M=20 (an M-shell is synonymous with L-shell for a dipole field but includes the contribution of the current sheet at Jupiter) is ~3.7 km/s, and the drift period would be ~60 hours. For ~100 eV energy particles, or at smaller M-shells, the drift would be even slower. In Saturn's magnetosphere, energy-banded ion distributions in the magnetodisc were observed by *Cassini* between Saturn L-shells of L=5.5-6.9 at 10-200 eV/q energies (Thomsen et al. 2017). Bounce resonance of the particles with standing Alfvén waves in the magnetosphere was proposed to be the explanation for the organized banding.

At Jupiter, banded protons and electrons were discovered in a narrow range of magnetic flux tubes connected to the orbits of the Galilean moons, at energies ranging from 0.2-46 keV (Sarkango et al., 2024). These moon-related flux enhancements were linked to bounce-resonance with standing Alfvén waves in the moons' orbits. At Jupiter, bounce frequencies for 0.1-1 keV ions are in the mHz range, which is comparable to eigenfrequencies of field-line resonances or standing Alfvén waves in the magnetosphere (Lysak & Song, 2020; Manners et al., 2018; Manners & Masters, 2019). Hence, bounce-resonance between these lower energy ions and field-line resonances was feasible. However, as noted in the discussion, the moon-related features and those presented in this work differ in various respects (see Section 3.4 and Table S3 in the Supporting Information).

This manuscript reports on banded energy distributions observed in Jupiter's magnetosphere by the JADE instrument onboard the *Juno* spacecraft at typical energies <10 keV, mapping to M-shells between M=10-20. In Section 2, we show the relevant observations of banded energy distributions. In Section 3, we discuss the interpretation and significance of the observations and summarize our findings in Section 4.

## 2 Observations

We used data collected by the Jovian Auroral Distributions Experiment ion sensor (JADE-I) onboard *Juno* near the polar regions of Jupiter (McComas et al. 2017). *Juno* orbits Jupiter in a highly elliptical trajectory and transits across a range of M-shells while in this region near perijove. During this time, JADE-I measured ion fluxes between energies of ~0.01-46 keV/q and mass-per-charge ($m/q$) of 1 to 64 using an electrostatic analyzer and a time-of-flight (TOF) section. The TOF data is usually averaged to longer cadence of at least one *Juno* spin period of ~30 s and therefore does not contain directional information about the incoming ions. Note that the data was not corrected for the Juno's velocity, i.e. the Compton-Getting effect, which may translate to an energy error of about ~10 eV for $H^+$ to ~100 eV for $O^+$ near the polar regions.

Since *Juno's* first Jovian close approach, heavy ions have been observed at high latitudes and their energy dependence was initially suggested to be related to their corresponding equatorial corotational energy (Szalay et al., 2017). *Juno* has since encountered numerous intervals near perijove in which banded energy distributions were present in different ion species (identified in $m/q$ vs. energy space) as measured by JADE, often simultaneously. We manually identified periods in the JADE-I data which exhibited multiple bands up to and including *Juno's* 63rd orbit (see Supporting Information for a list). In Figure 1a, we show the locations of the identified ion banding events in a magnetic coordinate system aligned with a dipole magnetic field, including the CON2020 current sheet (Connerney et al., 2020), to illustrate where these events map to in the magnetodisc. The dipole approximation allows us to compare the magnetic latitude of each event and condense events across longitude, which would otherwise not be possible if using a non-axisymmetric field. Figure 1b-1d shows an example of ion banding during *Juno*'s 61st orbit. In contrast to the banded distributions associated with the Galilean moons (Sarkango et al., 2024) that are shown in red, these new observations (purple) map to M-shells M>10 and were observed at higher altitudes ($h > 0.5$ $R_J$).

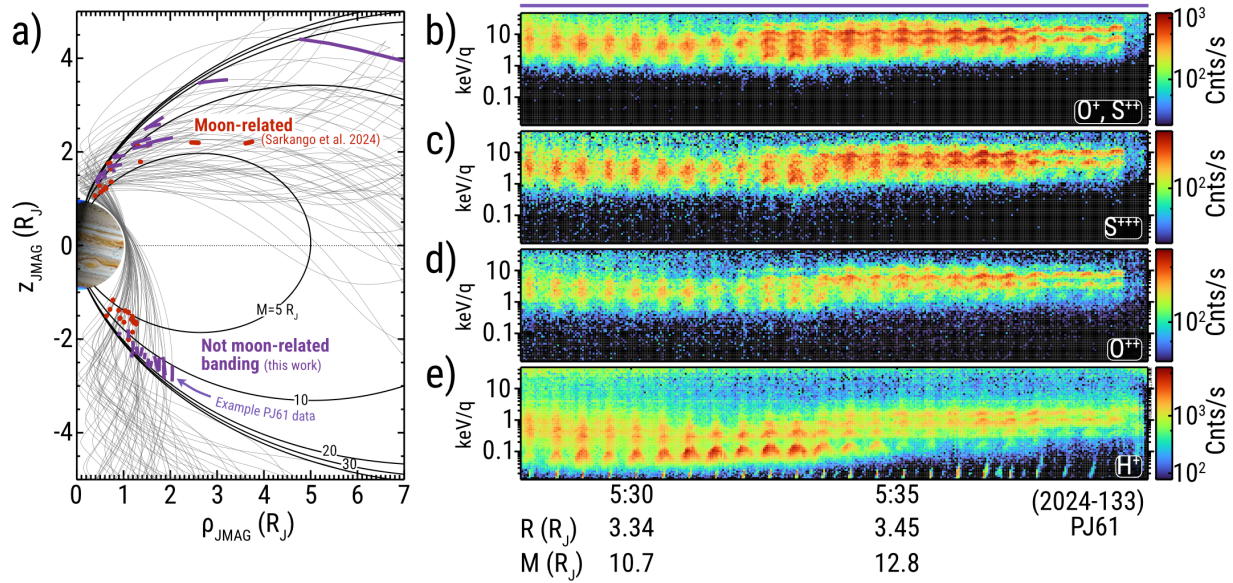

**Figure 1.** (a) Locations of all possible banded energy distributions observed in the JADE data so far between 2016-2024 in a dipole-centric coordinate system. Moon-related intervals discussed in Sarkango et al. (2024) are highlighted in red. Other intervals with ion banding discussed in the present work are highlighted in purple. (b-e) Energy distributions during the southern pass near perijove 61 (2024, DOY 133) of different ion species ($H^+$, $O^{++}$, $S^{+++}$, $O^+/S^{++}$) separated by JADE-I using the TOF measurement.

We show one such event in more detail (other events are also analyzed in the Supporting Information). Figures 1b-1e show the high-rate energy spectra for different mass-per-charge ion species as measured by JADE. These spectrograms are created by converting from TOF x E to (ion mass in AMU) x E space using a response function (Kim et al., 2020), then averaging over the count rates in AMU x E space within a window of ±10% of each $m/q$ value, following previous techniques for generating species-dependent energy-time spectrograms (Szalay et al., 2024). Figures 1b-1e focus on a period of ~10 min during PJ 61 (2024, DOY 133). Note that this interval is much longer than those found associated with the moon footprint tail crossings (Sarkango et al., 2024), which lasted for ~10-60 s near the moons' auroral footprints. During this longer interval, *Juno* was transiting across M-shells from approximately M=10 to 20, i.e. mapping to larger M-shells with increasing time. Flux enhancements (i.e., bands) were observed for $H^+$ between ~0.1-2 keV, and for heavier species ($O^{++}$, $S^{+++}$, and $O^+/S^{++}$) between ~1-20 keV. The 30 s intermittency in the bands is due to *Juno*'s spin causing a regular time-varying look direction dependence. This indicates a pitch-angle dependence of the bands because of JADE-I's changing field-of-view. For all ion species, the band at the lowest energies (~0.1 keV for H$^+$) was absent for M>13. The disappearance of lowest-energy bands with increasing M-shell was also seen in the *Cassini* H$^+$ observations by Thomsen et al., (2017).

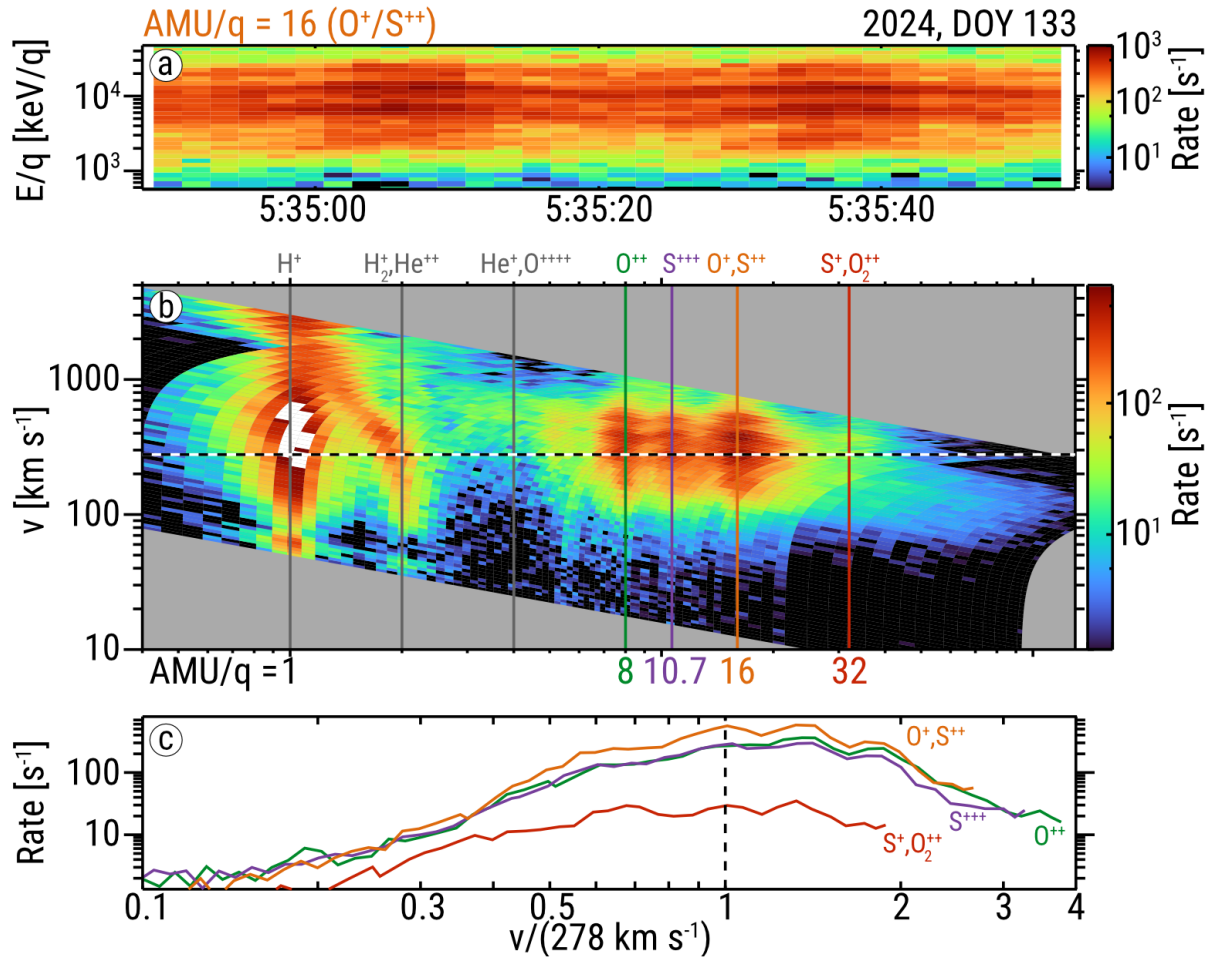

**Figure 2.** *Processed time-of-flight data from JADE-I during the interval shown in Figure 1, showing (a) the energy spectra extracted for m/q=16 ions, (b) the count-rate at different measured particle speeds and mass-per-charge (m/q). Different ion species can be identified based on their m/q ratios. c) Measured count-rate extracted along specific m/q ratios corresponding to the major heavy ion species. Banded flux intensifications are present at nearly the same speeds for all measured heavy ion species. The dashed line is centered on a band at v =278 km s$^{-1}$, the approximate location of the first peak.*

Figure 2 shows a processed JADE-I time-of-flight spectra during the event in Figure 1. Panel a) shows the energy spectra extracted for $m/q$=16, and panel (b) shows the measured count rates with ion mass-per-charge on the X-axis and particle speed on the Y-axis. This representation of the rates demonstrates that although the bands were observed at different energies for different ion species, they occur at the same speeds across all species. This can also be seen visually in Figure 2c, where the count-rate spectrum was extracted along each fixed $m/q$ value.

Since particle bounce periods depend on their speed, these different ion species which exhibit flux enhancements at particular speeds would have the same bounce periods for a given equatorial pitch angle. Since these observations were made at high-latitudes, JADE is sampling only a fraction of the trapped population that have low equatorial pitch angles (typically < 2°) and are able to reach these latitudes. For these small equatorial pitch angles, the bounce periods are more sensitive to energy than the absolute value of the equatorial pitch angle.

We *approximate* bounce periods by the following expression (Baumjohann & Treumann, 2022),

$$\tau_b = 4 \int_0^{s_m} \frac{ds}{v_\parallel} = 4 \int_0^{s_m} \frac{ds}{v \sin \alpha} = \frac{4}{v} \int_0^{s_m} \frac{ds}{\sqrt{1 - \frac{\sin^2 \alpha_{eq}}{B_{eq}} B(s)}}.$$

(1)

where $v$ is the particle's speed, $B_{eq}$ is the equatorial field strength at a particular M-shell, $\alpha$ is the local pitch angle, $\alpha_{eq}$ is the equatorial pitch angle for the particle, and $B(s)$ is the field strength at a parameterized distance $s$ along the field line. More details are provided in the Supporting Information. This simplified approximation represents the "average" bounce period across all longitudes and is sufficient for the qualitative comparisons shown in this manuscript. Bounce frequencies in [Hz] are the inverse of the bounce period ($\omega_b = 1/\tau_b$) in [s].

## 3 Interpretation

### 3.1. Occurrence of the banded distributions

The banded ions have been consistently observed throughout *Juno*'s tour through the Jovian magnetosphere (see Figures S1-S33). There is significantly more organization to when/where these bands are observed in the northern vs. southern hemisphere. *Juno* persistently observed banded ions in the southern hemisphere when it transited radial distances of 2.1-3.5 $R_J$ and M-shells of M=9-22. However, in the northern hemisphere, these features were observed in a more extended region and were less organized in M-shell. Some of this discrepancy may be due to the observational bias and asymmetry in coverage of *Juno*'s orbit between the two hemispheres. Alternatively, this could be a result of a more dipolar southern hemisphere and a highly non-axisymmetric magnetic field in the northern hemisphere that introduces azimuthal asymmetry (Connerney et al., 2018). Yet, the persistent observation of these bands provides evidence that the process of sustaining them is a fundamental process operating continually in Jupiter's magnetosphere.

Banded distributions were observed near the polar regions, mapping to M~10-20. *Juno* has transited through the equatorial regions between the same M-shells, but banded energy distributions were not observed by JADE near the equator. Perhaps this is because bounce periods of particles are dependent on equatorial pitch angle. Near the equator, particles are observed for a range of equatorial pitch angles, and there are many combinations of pitch angles and energies that would be bounce-resonant at some frequency. In contrast, observations made in the polar regions near Jupiter on similar M-shells represent a narrow subset of the trapped population (typically $\alpha_{eq} < 2°$), and the bounce periods for this subset population are more sensitive to particle energy than equatorial pitch angle. Hence, this is one possibility for why flux enhancements in energy, representative of bounce-resonance, are so prominent near the poles. This hypothesis is consistent with the bands observed during PJ 61, which were intermittent at Juno's spin period. Especially at the band at the lowest energy (Figure 1d), a mini-dispersion with time is clearly visible at the 30-s period, likely due to JADE sampling different pitch-angles during *Juno*'s spin. However, this dispersion does not appear prominent at higher energies. Note that JADE's energy channels are spaced logarithmically, so the bandwidth of the highest energy channels (e.g. at 10 keV) is much larger than the bandwidth of the smaller channels (e.g. at 10 eV), and small dispersions (e.g. at $\Delta E$=10 eV) are hence resolved insufficiently at the highest energies.

It is also worth asking why the banded distributions were only observed between M=10-20. Beyond M=20, plasma is not rigidly corotating, and there exists a strong current sheet. The presence of a current sheet could interfere with the bounce motion (Cheng & Decker, 1992; Speiser et al., 2013), although this effect is weaker for lower energy particles whose gyroradii are typically much smaller than the radius of magnetic curvature. We also note that it is unlikely that the observed bands result from instrumental artifacts, since the banded distributions were observed in the same polar regions across many perijove passes and were observed for different ion species at the same speeds.

### 3.2. Correspondence between bounce-periods and the System-III periodicity

In Figure 3a we use the example from PJ 61-S (from Figure 1) to highlight the different bands in the H$^+$ spectra. Points were identified for each band by eye and are highlighted in distinct colors in panel (a). Each band has some bandwidth that is most prominently seen in the lowest energy band that ranges from ~30-100 eV. Other bands also have a finite width in energy, but this is difficult to resolve at the higher energy levels, since the bands are resolved in fewer energy channels.

The bounce frequencies for particles depend on the field line geometry and vary with M-shell, in addition to equatorial pitch angle and energy. The corresponding bounce frequencies for each band in Figure 3a, calculated using a JRM09-dipole and CON2020 current-sheet model and assuming a 3° equatorial pitch angle (choosing 1° instead would not particularly change the results) are shown in Figure 3b and they exhibit the linear spacing that is characteristic of bounce resonance. Error-bars are shown for the lowest energy band in red representing the energy bandwidth (identified by eye), but the same is not done for the higher energies as the energy resolution gets worse at higher energies.

Jupiter rotates at the System-III frequency $(\Omega_J = 1/(9.92 \text{ hours}) = 0.028 \text{ mHz})$ (Seidelmann & Divine, 1977) and the magnetospheric plasma does not corotate rigidly beyond M~15-20 (Bagenal et al., 2016), so the flow sub-corotation frequency $\omega_\phi$ is smaller than the System-III frequency ($\omega_\phi < \Omega_J$). It can be seen in Figure 3b that bounce frequencies of different bands show good correspondence with integer multiples of the System-III frequency $(\Omega_J, 2\Omega_J, 3\Omega_J, ...)$ shown as horizontal dashed lines. This is indicative of a resonant interaction between the mirroring plasma ions at energies <10 keV and rigid-corotation with Jupiter, which has not been observed or discussed before in the literature.

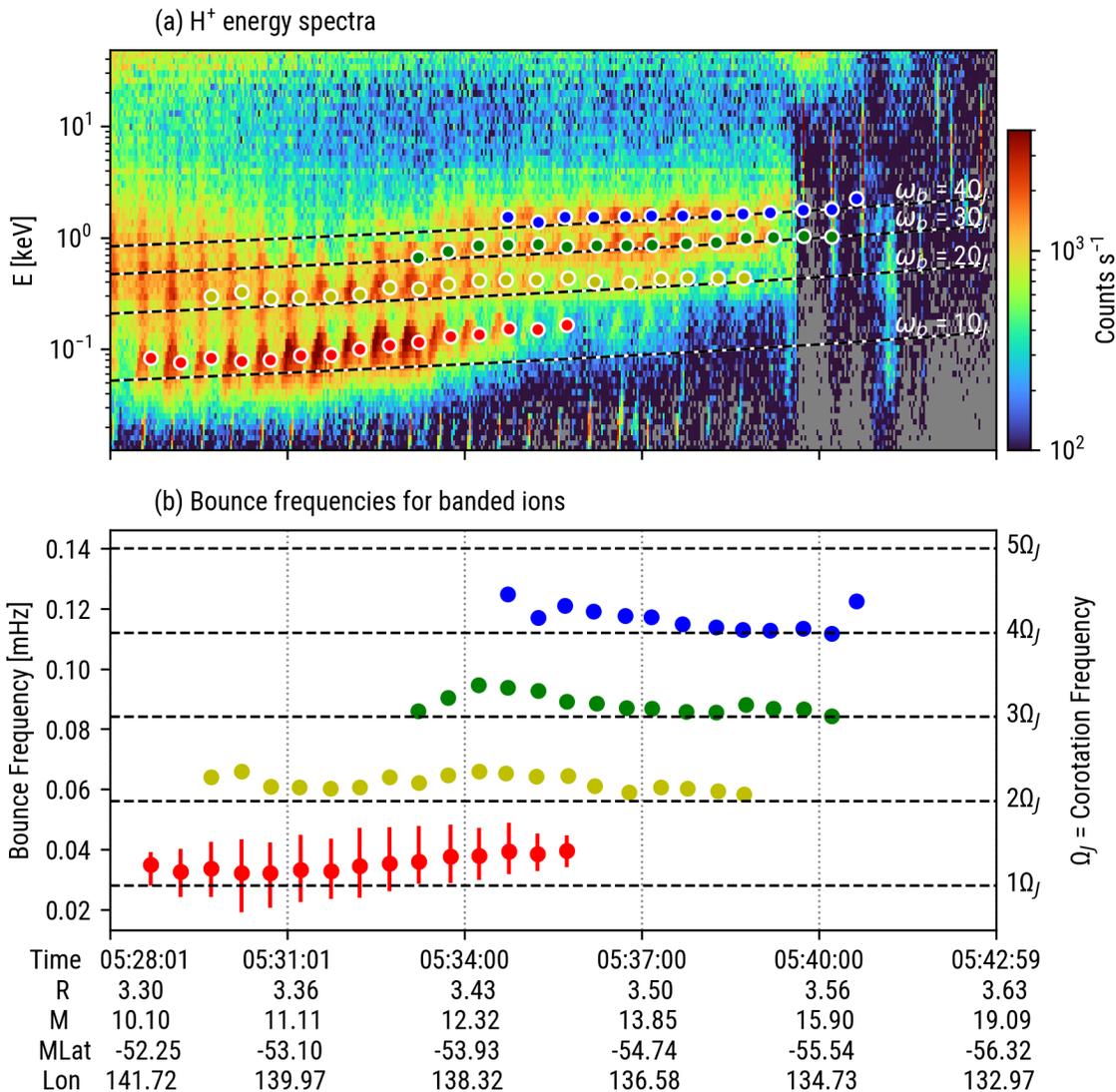

**Figure 3.** *(a) H⁺ energy-time spectra during the same interval as in Figure 1 with different bands highlighted in red, yellow, green, and blue points. (b) Bounce frequencies for the banded ions (assuming $\alpha_{eq} = 3°$) compared to integer multiples of the corotation or System-III frequency with respect to Jupiter.*

### 3.3. Hypotheses for corotation-bounce resonance

How would particles that are trapped and bouncing along magnetic field lines experience a resonant interaction with corotation? In particular, resonant wave-interaction requires a modulated electric field in the frame of the particle. We present a few hypotheses for consideration.

1. **Bounce-resonance with Alfvén waves:** Similar to Thomsen et al., (2017)'s Saturn observations, it is possible that the banding results from bounce-resonance within standing Alfvén waves. Eigenperiods for field-line resonances between M=10-20 are on the order of 10-80 min (only the fundamental is larger than 80 min) (Lysak & Song, 2020). Meanwhile, harmonics of the corotation period are at ~595 min, 297 min, 198 min, 148 min, 119 min, …, which coincidentally are also the bounce periods for the low-energy ions within the bands. However, the correspondence between corotation and the bounce frequencies of the banded ions cannot be explained easily. One possibility is that corotation acts as a doppler-shift in the drift-bounce resonance condition. That is, resonance occurs when $\omega - m\Omega_J = N\omega_b$, where $\Omega_J$ is the rotation frequency of Jupiter. However, in this model, the Alfvén wave has to be "standing" in the non-rotating frame. It is unclear how this is consistent with the fact that Alfvén waves travel predominantly along magnetic field lines and are frozen-in with the corotating plasma, i.e., the field-lines are also corotating. Conversely, it is also possible that the observations reported by Thomsen et al. (2017) resulted from a similar corotation-bounce resonance (see Figure 4 for the corotation bounce-resonant energy at Saturn that falls within the CAPS energy range), and it may be useful to revisit the CAPS data.

2. **Bounce-resonance with a dawn-dusk electric field:** Another way for the particle to perceive an electric field modulation is if it encounters a spatially varying electric field structure during its corotation $\boldsymbol{E} \times \boldsymbol{B}$ drift, e.g., the dawn-dusk electric field at Jupiter. A resonant interaction is possible if the bounce frequency of the particle corresponds to the corotation frequency. This situation is analogous to the model of bounce-drift resonance within standing Alfvén waves. The resonance condition is a special case of bounce-drift resonance with $\omega = 0$, i.e., $m\omega_d = N\omega_b$, where $\omega_d$ is the frequency at which the particle is drifting azimuthally through the electric field structure ($\omega_d = \Omega_J$ for corotating particles). However, plasma between M=10-20 is more likely sub-corotating (e.g. $\omega_d \sim 0.7\Omega_J$) (Bagenal et al., 2016), though recent Juno observations also show strongly corotating plasma (Kim et al. 2020; J. Wang et al. 2024). The observations during PJ61 appear to be consistent with bounce-resonance with the System-III/rigid-corotation periodicity ($\Omega_J$), but there are other events (among Figures S1-S35) that follow or lie between, but do not exactly match the System-III harmonics.

3. **Bounce-resonance due to explicit time-variability of the background magnetic field:** Jupiter's magnetodisc wobbles due to the non-axisymmetric magnetic field of Jupiter at the System-III periodicity (Krupp et al., 2004). This periodicity is transmitted to the magnetosphere via Alfven waves that are associated with a non-zero $\partial \boldsymbol{B}/\partial t$, i.e., an induced electric field. Particles may resonate with this periodic perturbation over their bounce motion.

Alternatively, Hones & Bergeson (1965) showed that in a tilted-dipole geometry, the $\boldsymbol{E} \times \boldsymbol{B}$ drift alone is not sufficient to drive stationary particles to corotation. They proposed that corotation could be achieved nevertheless via Fermi acceleration with the changing background magnetic field. This process could occur in a resonant manner.

### 3.4. Comparison with moon-related banding events in Sarkango et al. (2024)

Note that the banded energy distributions observed in Sarkango et al. (2024) mapped to a narrow range of M-shells of a few moon-radii across the moons' orbit, and were observed at ~0.2-1 keV for electrons and ~>1 keV for protons, but never simultaneously for both species. In contrast, the events discussed in this work lasted for ~10 min, mapped to a large range of M-shells (M=10-20), and were seen simultaneously for multiple ion species, at energies <1 keV. In fact, the moon events are embedded within this larger banded structure (see Figure S11 in the Supporting Information). We tabulate these differences in more detail in Table S3 in the Supporting Information.

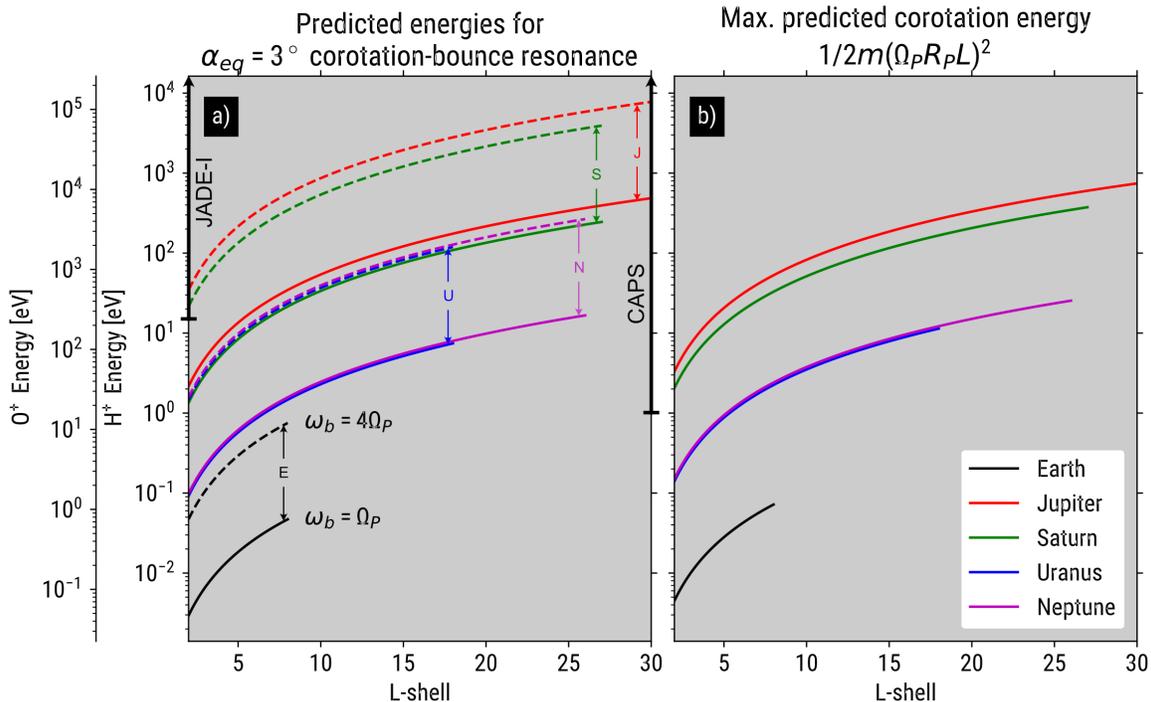

***Figure 4.*** *(a) Energies at which H$^+$ and O$^+$ ions with $\alpha_{eq} = 3°$ would be expected to be in $\omega_b = \Omega_P$ and $\omega_b = 4\Omega_P$ resonance, where $\Omega_P$ is the planetary rotation frequency for Earth, Jupiter, Saturn, Uranus, and Neptune, marked in different colors. (b) The kinetic energy of the ion associated with corotation flow (i.e. $mv_{cor}^2/2$) where $v_{cor} = \Omega_P L R_P$ for a given L-shell and planetary radius $R_P$. We use L-shell here because we have assumed a dipole field for all planets for this preliminary and qualitative comparison. Note that while the curves shown in panels (a) and (b) look similar, they are not expected to be related and differ by a factor of 1.4.*

### 3.5. Corotation-bounce resonance at other planets

Assuming a dipole field, we can roughly parameterize the relevance of corotation-bounce resonance in different planetary magnetospheres, namely those of Earth, Jupiter, Saturn, Uranus, and Neptune. In Figure 4a we show the energies at which H$^+$ and O$^+$ ions would be expected to be in $\omega_b = \Omega_P$ and $\omega_b = 4\Omega_P$ resonance at the planet's rotation frequency $\Omega_P$. Figure 4b shows the maximum corotation energy ($mv_{cor}^2/2$) expected at a particular L-shell. Although the two curves in panels (a) and (b) have similar trends (both happen to scale as $E \propto \Omega_P^2 R_P^2 L^2$), they are not related. The equatorial corotation energy is 40% larger than the $\omega_b = \Omega_P$ resonant energy. Figure 4a clearly shows that bounce-resonance with corotation can occur for H$^+$ at ~1 eV-5 keV at Jupiter and Saturn, at 0.1 eV-100 eV at Uranus and Neptune, and is not expected to occur at any energies >1 eV for the terrestrial magnetosphere. The range of possible resonant energies at Jupiter and Saturn is well within the detection capabilities of *Juno*'s JADE and *Cassini's* CAPS instruments. Figure 4 demonstrates that corotation-bounce resonance would be possible at very different energies in the planets considered due to different bounce periods and corotation energies at each body. While we have included Jupiter's current sheet in the magnetic field line geometries for the rest of the analysis in this study, we have ignored the current sheet in Figure 4 to provide consistent comparisons across the planets (hence the use of L-shell on the X-axis). We found the expected resonant energies are less than a factor of 2 different at Jupiter (not shown) by considering the current sheet and therefore it does not significantly affect this calculation.

## 4 Summary

Simultaneous banded energy distributions of different ion species in Jupiter's polar regions mapping to M-shells between roughly M=10-20 were often observed by *Juno*'s JADE instrument. The bands were observed at different energies but similar speed for different ions, hinting at a bounce-resonant interaction. The bounce frequencies of ions at the banded energies usually corresponded to integer multiples of the System-III or rigid-corotation frequency ($\Omega_J$) of Jupiter. We have shown one event in detail in this manuscript and presented analysis of several events (N=33) in the Supporting Information.

A bounce resonant wave-particle interaction is possible if the particle perceives an electric field modulation to resonate with during its bounce motion. Corotation-bounce resonance could occur via Alfvén waves, or with a spatially varying electric field that the particle drifts in and out of, or with an explicitly time-varying electric field such as an induced electric field produced by $\partial \boldsymbol{B}/\partial t$ of a changing magnetic field of a tilted dipole. Beyond mentioning these non-exhaustive possibilities, we do not further speculate.

These novel observations illustrate an unexpected link between bounce motion and plasma corotation at low energies in Jupiter's magnetosphere, i.e., a corotation-bounce resonance (CBR) that is different from other resonances discussed previously in literature. These distributions have been observed frequently by *Juno* and mapping to the same range of M-shells in the middle magnetosphere, indicating that the as-of-yet unknown resonant mechanism that produces these distributions is a fundamental and important process operating in Jupiter's magnetosphere. Corotation-bounce resonance may also be occurring at similar energies in Saturn's magnetosphere, and encourages a revisit of the Thomsen et al. (2017) observations. If operating also at Uranus and Neptune, corotation-bounce resonance would be relevant for $H^+$ ions at energies $<\sim100$ eV and $O^+$ ions at energies $<\sim1$ keV. This process may not be relevant for the terrestrial magnetosphere due to the much shorter bounce periods for plasmaspheric ions and electrons compared to Earth's 24-hour rotation period.

### Acknowledgments

We are grateful for the efforts of the *Juno* mission teams, allowing for unique opportunities to study the Jovian magnetosphere. This work was supported by NASA grants NFDAP 80NSSC23K0276 and NFDAP 80NSSC23K0665, and through the *Juno* mission's JADE instrument science team. J. Saur acknowledges funding from the Deutsche Forschungsgemeinschaft (SA 1772/6-1). Y. S. is grateful to V. Dols, Y.-X. Hao, and Z.-Y. Liu for insightful discussions.

### Open Research

All JADE data analyzed in this work is publicly available from the NASA Planetary Data System Plasma Interactions Node and can be found at Allegrini, F. et al., (2024).

*Geophysical Research Letters*

Supporting Information for

# Corotation-bounce resonance of ions in Jupiter's magnetosphere

Y. Sarkango[1], J. R. Szalay[1], P. A. Damiano[2], A. H. Sulaiman[3], P. A. Delamere[2], J. Saur[4], D. J. McComas[1], R. W. Ebert[5,6], F. Allegrini[5,6]

[1]Department of Astrophysical Sciences, Princeton University, USA

[2]Geophysical Institute, University of Alaska Fairbanks, USA

[3]School of Physics and Astronomy, University of Minnesota, USA

[4]Institut für Geophysik und Meteorologie, Universität zu Köln, Germany

[5]Southwest Research Institute, USA

[6]Department of Physics and Astronomy, University of Texas at San Antonio, USA

Corresponding author: Yash Sarkango (sarkango@princeton.edu)

***Contents of this file***

**Table S1. List of non-moon-related banding intervals.***

| PJ | Start | Stop | Start M | Stop M | Start r [R_J] | Stop r[R_J] |
|---|---|---|---|---|---|---|
| 1 | 2016-240T04:36:41 | 2016-240T08:17:52 | 15.08 | 16.94 | 9.94 | 6.48 |
| 1 | 2016-240T12:12:06 | 2016-240T12:15:15 | 16.67 | 9.58 | 1.64 | 1.57 |
| 7 | 2017-191T22:53:54 | 2017-191T23:16:58 | 13.2 | 20.99 | 4.79 | 4.33 |
| 7 | 2017-192T01:19:10 | 2017-192T01:20:32 | 13.05 | 10.17 | 1.57 | 1.54 |
| 17 | 2018-355T16:17:15 | 2018-355T16:32:45 | 14.76 | 18.79 | 1.73 | 1.39 |
| 18 | 2019-043T16:43:42 | 2019-043T16:49:20 | 10.11 | 18.34 | 1.93 | 1.79 |
| 19 | 2019-096T10:24:06 | 2019-096T10:44:06 | 11.69 | 11.8 | 3.31 | 2.86 |
| 20 | 2019-149T06:24:46 | 2019-149T06:35:00 | 13.09 | 24.88 | 3.15 | 2.92 |
| 21 | 2019-202T03:00:50 | 2019-202T03:07:59 | 10.68 | 36.33 | 2.19 | 2.02 |
| 22 | 2019-255T02:25:58 | 2019-255T02:28:13 | 13.42 | 16.06 | 2.49 | 2.44 |
| 27 | 2020-154T08:57:32 | 2020-154T09:10:19 | 8.63 | 15.31 | 2.69 | 2.38 |
| 32 | 2021-052T16:13:46 | 2021-052T16:26:12 | 11.07 | 17.48 | 2.79 | 2.49 |
| 33 | 2021-105T22:41:16 | 2021-105T22:44:21 | 12.11 | 20.59 | 1.94 | 1.86 |
| 34 | 2021-159T06:57:57 | 2021-159T06:59:04 | 13.37 | 16.09 | 1.86 | 1.83 |
| 35 | 2021-202T07:21:51 | 2021-202T07:26:53 | 9.69 | 18.07 | 1.98 | 1.86 |
| 37 | 2021-289T15:33:47 | 2021-289T15:53:24 | 7.77 | 9.5 | 3.09 | 2.63 |
| 37 | 2021-289T16:13:45 | 2021-289T16:18:33 | 16.27 | 25.9 | 2.14 | 2.03 |
| 37 | 2021-289T18:11:47 | 2021-289T18:19:09 | 10.25 | 15.22 | 2.1 | 2.28 |
| 38 | 2021-333T13:34:27 | 2021-333T13:45:09 | 15.34 | 17.04 | 1.64 | 1.4 |
| 38 | 2021-333T15:29:31 | 2021-333T15:34:20 | 13.14 | 17.89 | 2.53 | 2.64 |
| 40 | 2022-056T02:56:49 | 2022-056T02:59:10 | 10.27 | 13.16 | 2.07 | 2.12 |
| 41 | 2022-099T16:58:26 | 2022-099T17:05:39 | 12.99 | 22.35 | 2.36 | 2.54 |
| 42 | 2022-143T03:31:08 | 2022-143T03:37:30 | 13.43 | 21.19 | 2.51 | 2.66 |
| 43 | 2022-186T08:31:31 | 2022-186T08:33:34 | 10.84 | 14.31 | 1.81 | 1.76 |
| 44 | 2022-229T16:09:47 | 2022-229T16:14:22 | 11.89 | 16.14 | 2.72 | 2.83 |
| 48 | 2023-022T07:02:58 | 2023-022T07:08:57 | 10.58 | 16.24 | 2.59 | 2.73 |
| 49 | 2023-060T07:23:34 | 2023-060T07:29:03 | 13.24 | 21.41 | 2.86 | 2.99 |
| 53 | 2023-212T10:40:54 | 2023-212T10:50:59 | 9.93 | 15.19 | 2.98 | 3.21 |
| 54 | 2023-250T13:31:35 | 2023-250T13:36:36 | 13.02 | 19.03 | 2.94 | 3.05 |
| 55 | 2023-288T12:36:01 | 2023-288T12:47:31 | 11.05 | 18.33 | 3.16 | 3.41 |
| 57 | 2023-364T14:13:04 | 2023-364T14:28:13 | 9.05 | 14.56 | 3.01 | 3.35 |
| 58 | 2024-034T23:04:37 | 2024-034T23:09:22 | 12.25 | 20.08 | 2.55 | 2.66 |
| 60 | 2024-100T10:19:34 | 2024-100T10:33:45 | 11.42 | 17.72 | 2.88 | 3.21 |
| 61 | 2024-133T05:27:19 | 2024-133T05:39:19 | 9.89 | 15.37 | 3.28 | 3.55 |
| 63 | 2024-198T16:03:33 | 2024-198T16:12:54 | 10.27 | 18.78 | 2.85 | 3.06 |

*See Figures S1-S33 for the ion energy spectra during these intervals.

**Table S2. List of moon-related banding intervals (Sarkango et al., 2024).**

| Moon | Start | Stop |
|------|-------|------|
| **Io** | 2017-086T09:30:51 | 2017-086T09:31:02 |
| **Io** | 2017-139T06:39:53 | 2017-139T06:40:04 |
| **Io** | 2017-192T02:22:27 | 2017-192T02:22:54 |
| **Io** | 2018-144T05:13:11 | 2018-144T05:13:20 |
| **Io** | 2018-197T04:49:50 | 2018-197T04:50:12 |
| **Io** | 2018-250T01:46:00 | 2018-250T01:46:17 |
| **Io** | 2019-043T14:55:50 | 2019-043T15:02:38 |
| **Io** | 2019-255T03:18:52 | 2019-255T03:19:07 |
| **Io** | 2019-307T22:07:02 | 2019-307T22:07:19 |
| **Io** | 2019-307T23:10:38 | 2019-307T23:10:56 |
| **Io** | 2019-360T18:30:16 | 2019-360T18:32:24 |
| **Io** | 2020-154T08:26:12 | 2020-154T08:31:19 |
| **Io** | 2021-052T18:23:13 | 2021-052T18:23:47 |
| **Io** | 2021-106T00:18:06 | 2021-106T00:18:43 |
| **Io** | 2021-2020T7:10:50 | 2021-202T07:11:10 |
| **Io** | 2021-245T22:21:57 | 2021-245T22:22:16 |
| **Io** | 2021-289T17:59:23 | 2021-289T17:59:56 |
| **Io** | 2021-333T15:10:17 | 2021-333T15:12:10 |
| **Io** | 2022-056T01:46:10 | 2022-056T01:46:17 |
| **Io** | 2022-099T16:40:10 | 2022-099T16:41:52 |
| **Io** | 2022-272T16:36:53 | 2022-272T16:37:17 |
| **Europa** | 2017-297T16:27:20 | 2019-297T16:27:50 |
| **Europa** | 2020-154T09:00:15 | 2020-154T09:00:35 |
| **Europa** | 2021-052T18:31:21 | 2021-052T18:31:31 |
| **Europa** | 2022-099T16:51:36 | 2022-099T16:51:50 |
| **Europa** | 2018-250T00:44:50 | 2018-250T00:45:02 |
| **Europa** | 2018-250T01:50:06 | 2018-250T01:50:10 |
| **Europa** | 2022-012T09:47:30 | 2022-012T09:47:58 |
| **Ganymede** | 2021-289T18:19:17 | 2021-289T18:20:07 |
| **Ganymede** | 2022-099T14:59:16 | 2022-099T15:00:41 |

**Table S3. Differences between Sarkango et al., (2024)'s moon-banding events versus corotation-resonance banding events discussed in this work.**

| | Moon banding events (Sarkango et al. 2024) | Corotation-resonance events (present work) |
|---|---|---|
| Duration | 10s to 1 min | >1 min to ~10 min |
| Particle species | $e^-$ and $H^+$, but never simultaneously | Simultaneously in H+, $O^{++}$, $S^{+++}$, and $O^+/S^{++}$<br><br>Never observed nor expected to be observed for $e^-$ since $e^-$ bounce much faster, and such resonance would occur at very low electron energies. |
| Energies | Typically, 0.1-1 keV for $e^-$, and >1 keV for $H^+$. | <1 keV for $H^+$ and banded energies scale with mass for heavier species. |
| M-shell range | Narrow range of M-shells (e.g. M=5.9-6.0) connected to the orbits of the Galilean moons. | Broad M-shell range for every event e.g. from M=10 to M=20. |
| Bounce periods of particles within banded energy distribution | <10 min for electrons, <100 min for protons | At higher harmonics of the corotation frequency, i.e. at ~595 min, 297 min, 198 min, 148 min. |
| Resonating with harmonics of specific frequency? | Unclear | Bounce frequencies of ions match are close to the System-III frequency and its higher harmonics. |

**Text S1. Peak count-rate vs. peak phase space density.**

The measured count-rate $R$ [count s$^{-1}$] is related to the phase space density $f(v)$ [s$^3$ m$^{-6}$] at some particle speed $v$,

$$R \propto v^4 f(v)$$

Eq (S 1)

Hence, the expected speed for peak count-rate (when $\partial R / \partial v = 0$) is not the same as the peak in phase space density when $\partial f / \partial v = 0$. Count-rate is maximum when

$$\frac{\partial R}{\partial v} = \left( 4v^3 f + v^4 \frac{\partial f}{\partial v} \right) = 0.$$

I.e. when,

$$4f + v \frac{\partial f}{\partial v} = 0.$$

Eq (S 2)

Assuming a Maxwellian distribution, $f = C e^{-\frac{(v-v_0)^2}{v_{th}^2}}$ , where $T$ is the temperature and $v_0$ is the bulk speed, and $v_{th} = \sqrt{k_B T / m}$ is the thermal speed, we find that

$$\frac{\partial f}{\partial v} = -\frac{2(v-v_0)}{v_{th}^2} C e^{-\frac{(v-v_0)^2}{v_{th}^2}} = -\frac{2(v-v_0)}{v_{th}^2} f.$$

Eq (S 3)

Which implies that, for a drifting Maxwellian distribution, $f$ peaks when $v = v_0$. Substituting Equation S3 into Equation S2 leads to a quadratic equation whose roots are at the peak count-rate-speed.

$$4 + v \left[ -\frac{2(v-v_0)}{v_{th}^2} \right] = 0$$
$$v^2 - v v_0 - 2 v_{th}^2 = 0$$

Hence, if the phase space density peaks at a speed $v_0$, the measured count-rate would peak at a different speed $v_c$, where

$$v_c = \frac{1}{2} \left( v_0 \pm \sqrt{v_0^2 + 8 v_{th}^2} \right).$$

Eq (S 4)

The same expression is also provided in Szalay et al., (2020). In terms of energy, the peak-count-energy is then,

$$E_c = \frac{1}{2} m v_c^2 = \frac{m}{4} \left( v_0^2 + 4 v_{th}^2 \pm v_0 \sqrt{v_0^2 + 8 v_{th}^2} \right).$$

Or,

$$E_c = \frac{E_0}{2} + m v_{th}{}^2 \pm \frac{m v_0}{4} \sqrt{v_0^2 + 8 v_{th}{}^2}.$$

<div align="right">Eq (S 5)</div>

Where $E_0 = 1/2(mv_0^2)$ is the energy at which the maximum phase space density is expected. If the temperature is negligible, e.g. for a beam or for cold plasma ($v_{th} \approx 0$),

$$E_c \approx \frac{E_0}{2} \pm \frac{m v_0^2}{4} \approx E_0.$$

That is, for a very cold plasma or beam, the peak in count-rate and the peak in phase space density would occur at the same energy.

For a warm plasma, if the thermal speed is the same as the mean speed, $v_{th} = v_0$, then

$$v_c \approx \frac{1}{2}\left( v_0 + \sqrt{v_0^2 + 8 v_0{}^2} \right) \approx 2 v_0, \text{ and}$$

$$E_c = \frac{1}{2} m v_c^2 \approx 4 \left( \frac{1}{2} m v_0^2 \right) \approx 4 E_0.$$

In other words, if the temperature of the plasma is 10 eV, and the expected corotation energy ($1/2 m v_{cor}{}^2$) is also 10 eV, then the energy at which a particle detector collects the most counts would be 40 eV. Interpretation of measured count-rate spectra in hot plasmas should account for this.

In Figure S0 we show $H^+$ energy distributions during an interval between 12:00-13:20 on 2020, DOY 101 (PJ 26) during which no bands were observed, however the count-rate peaks at a certain energy. During this time *Juno* is in the polar regions of Jupiter at an M-shell of M=9.8, and transits through magnetic field lines until it reaches a maximum M-shell of M=24, and then returns to smaller M-shell values of M=7.7. Over this time, the peak-count energy increases until the maximum M-shell value and then decreases.

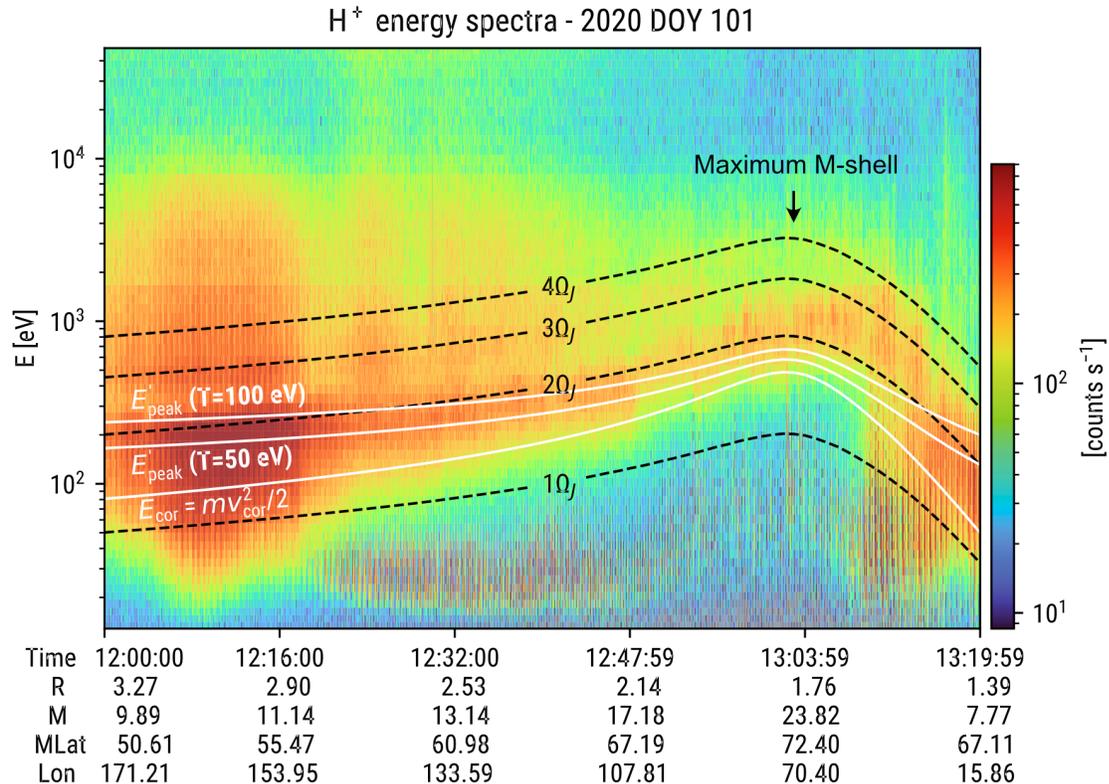

**Figure S0.** H+ energy spectra during an interval on 2020, DOY 101 (PJ 26). Except for a singular peak energy with maximum counts, no additional bands were observed during this interval. Dashed black curves indicate energies at which H+ ions are expected to be bounce-resonant with the corotation frequency $\boldsymbol{\Omega_J}$. The lowest white curve indicates $\boldsymbol{E_{cor} = \frac{1}{2} m_p v_{cor}{}^2}$, where $\boldsymbol{v_{cor} = \Omega_J \times r_{eq}}$, and $\boldsymbol{r_{eq}(M) = M}$, i.e., $\boldsymbol{E_{cor}}$ is the energy associated with proton corotation flow at the equator for some M-shell.

Assuming a Maxwellian distribution in phase-space that is corotating with Jupiter, the peak phase space density is expected to be at $E_{cor}$ where $E_{cor} = \frac{1}{2} m_p v_{cor}{}^2$ and $v_{cor} = \Omega_J \times r$ (here $\Omega_J$ needs to be in units of [rad s-1]), so it is expected that this energy also increases with increasing M-shell. We calculated $E_{cor}$ by identifying Juno's M-shell using the JRM09 internal field and the CON2020 current sheet model (Connerney et al., 2018, 2020). The blue curve in Figure 4 shows the expected $E_{cor}$ at the magnetic equator. The $E_{cor}$ curve follows the energy of the peak-count-rate but differs by a factor of ~2.

The energies at which particle bounce frequencies correspond to harmonics of the corotation frequencies, i.e., when $\omega_b = [\Omega_J, 2\Omega_J, 3\Omega_J, \dots]$ Hz, are shown as dashed black lines in Figure S0. In particular, there is close correspondence between peak-count energy and the energy at which $\omega_b = 2\Omega_J$. That is, even in the case when particle distributions are not banded, there appears to be some correspondence between particle bounce motion, the System-III periodicity, and the plasma corotation, since all three curves in Figure S0 – 1) energy with peak count-rate, 2) energy that is bounce-resonant with System-III, and 3) the energy associated with corotation flow, appear to vary in a similar manner.

### Text S2. Calculating M-shell and particle bounce periods

We approximate bounce periods by first identifying Juno's M-shell. The concept of M-shell is borrowed from the terrestrial community who define L-shell based on the radius of closed particle drift paths. At Jupiter, the strong current sheet distends the internal field of the planet, making a dipole L-shell an inadequate assumption. Hence, the outer planet community uses the concept of M-shell, which is roughly the equatorial distance of a particle corotation drifting in this non-dipolar magnetic field that includes the current sheet at all longitudes.

However, due to the presence of a magnetic field tilt, or non-axisymmetry (e.g. the JRM09 and JRM33 fields are highly non-axisymmetric), there is no way to uniquely characterize a field line using an M-shell value at Jupiter. Hence, there is no universally accepted definition for M-shell in Jupiter's magnetosphere, beyond the requirement that the current sheet field must be considered. In a non-axisymmetric magnetic field, the radial location at which the field intersects the equatorial plane varies with longitude. If so, should we then consider the average radial extent across all longitudes, or the maximum radial extent across all longitudes?

To circumvent these issues, we use the following simplified definition – an M-shell is the radial distance of the magnetic equator (where $B_r$ changes sign) of a field line. We trace the field line using the JRM09 field, including the CON2020 current sheet, and identify the point furthest to the planet, and use the radial distance as M-shell. This is not a robust calculation; it is a simple approximation. We use JRM09 instead of JRM33 because the current sheet fit parameters of Connerney et al., (2020) were obtained using the JRM09 field to fit the observations. The JRM33 field was produced after the CON2020 model. Hence, it is appropriate to use only the JRM09 field with the CON2020 model, even though the JRM33 field is more accurate near the planet. For a more detailed discussion on M-shell calculation see Rabia et al., (2024).

With this M-shell value, which is an approximation, we calculate the bounce period. One possibility is to trace the particle along the fixed field line and calculate how long it would take to return to the original location along that field line. However, in reality, for a particle with bounce period of ~9.92 hours, it is unrealistic to expect that the background field is constant. The motion of a particle in a time-varying field is not trivial (as perhaps the results in this manuscript illustrate). There is no guarantee that the second adiabatic invariant is conserved when the field changes faster than the bounce period.

This leads to the second approximation - we use the "dipole equivalent" of the JRM09 field. The JRM09 can be described as a tilted dipole. This approximation is worse closer to the planet, but reasonable far away from the planet as the higher order spherical harmonics decay faster than the lower order harmonics. With this "equivalent dipole", we center the Z-axis on the northern pole such that the magnetic equator is always in the XY plane. This is also the coordinate system used by Connerney et al., (2020) for Figure 1 in their manuscript where they defined the CON2020 coordinate system. All of Juno's dipole frame trajectory plots in the literature are typically defined in the same coordinate system that we use to calculate M-shell, using the "equivalent dipole" of the JRM09 (or VIP4) field. The advantage of this simplification is that now we do not have to consider any $\partial \boldsymbol{B}/\partial t$, since the field is axisymmetric about the dipole pole.

With this simplification, we calculate $\partial B/\partial s$ along the field line using second order central-differences and perform the numerical integration as shown in Equation 1 in the manuscript. The $v$ used in Equation 1 is the relativistic speed, though this correction is unnecessary for the lower energy ions used in this study. Our resulting bounce periods agree well with the established literature and to the observations shown in this work. We also find the approximate bounce periods track the observed corotation harmonics remarkably well (Figures S1-S35). For the qualitative discussion of this manuscript, such a simplified calculation is probably sufficient. Future work can consider the issues highlighted here in more depth using numerical models.

**Text S3. Description for Figures S1-S33**

In Figures S1-S33, we show the energy-time count-rate spectra for the ion species with m/q=1, 8, 32/3, and 16, representation the $H^+$, $O^{++}$, $S^{+++}$, and $O^+/S^{++}$ ion species that are commonly found in Jupiter's magnetosphere. Overlaid on each spectrum in dashed lines is the expected energy when the bounce periods of these ions matched harmonics of the System-III corotation frequency ($\Omega_J$, $2\Omega_J$, $3\Omega_J$, …). The lowest dashed line represents $\omega_b = \Omega_J$. We calculate approximate bounce periods by considering the M-shell of Juno at each measurement (as described above). Hence, the dashed curves are not constant in energy or time, and their variation reflects Juno's changing M-shell.

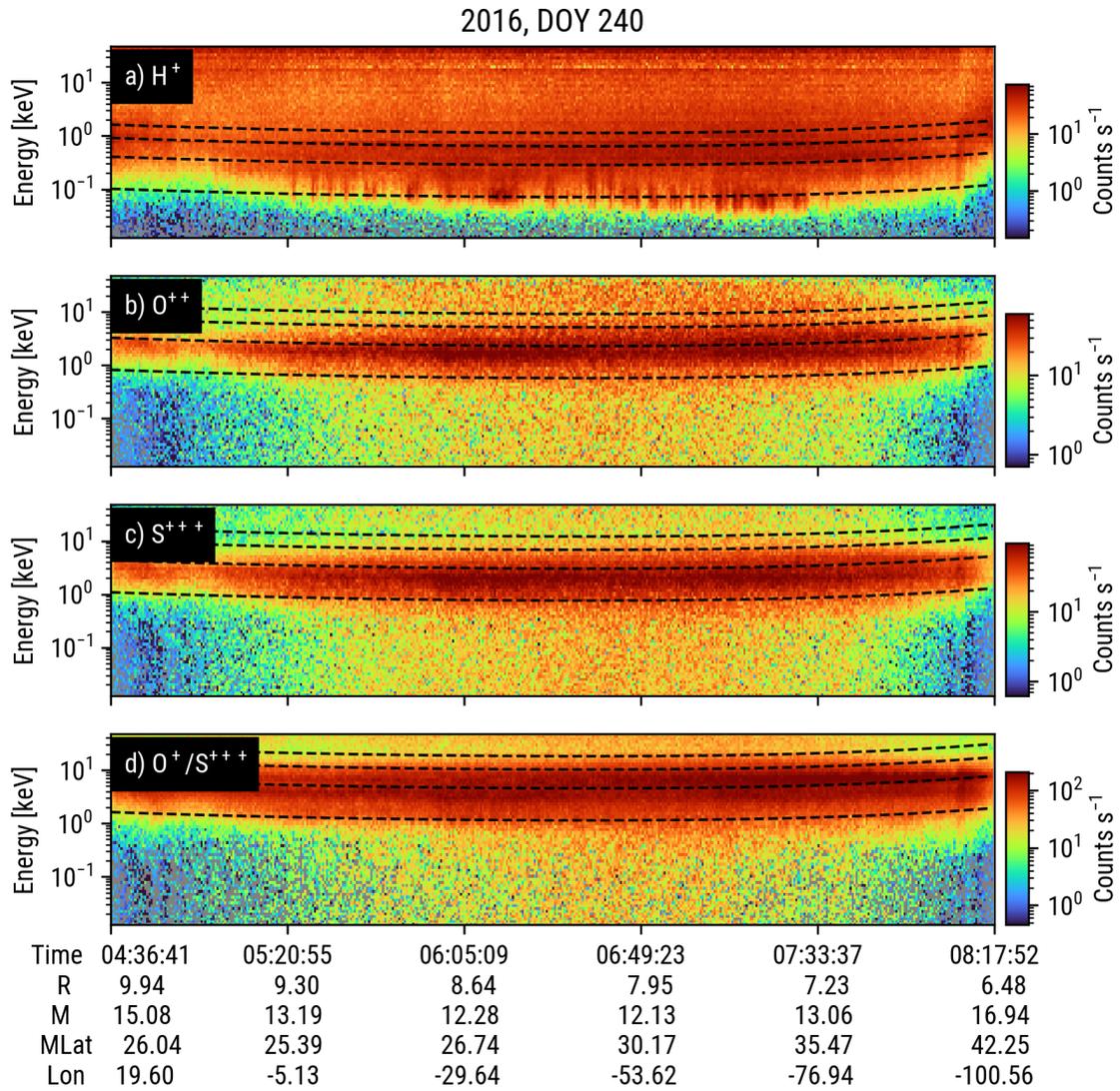

Figure S 1

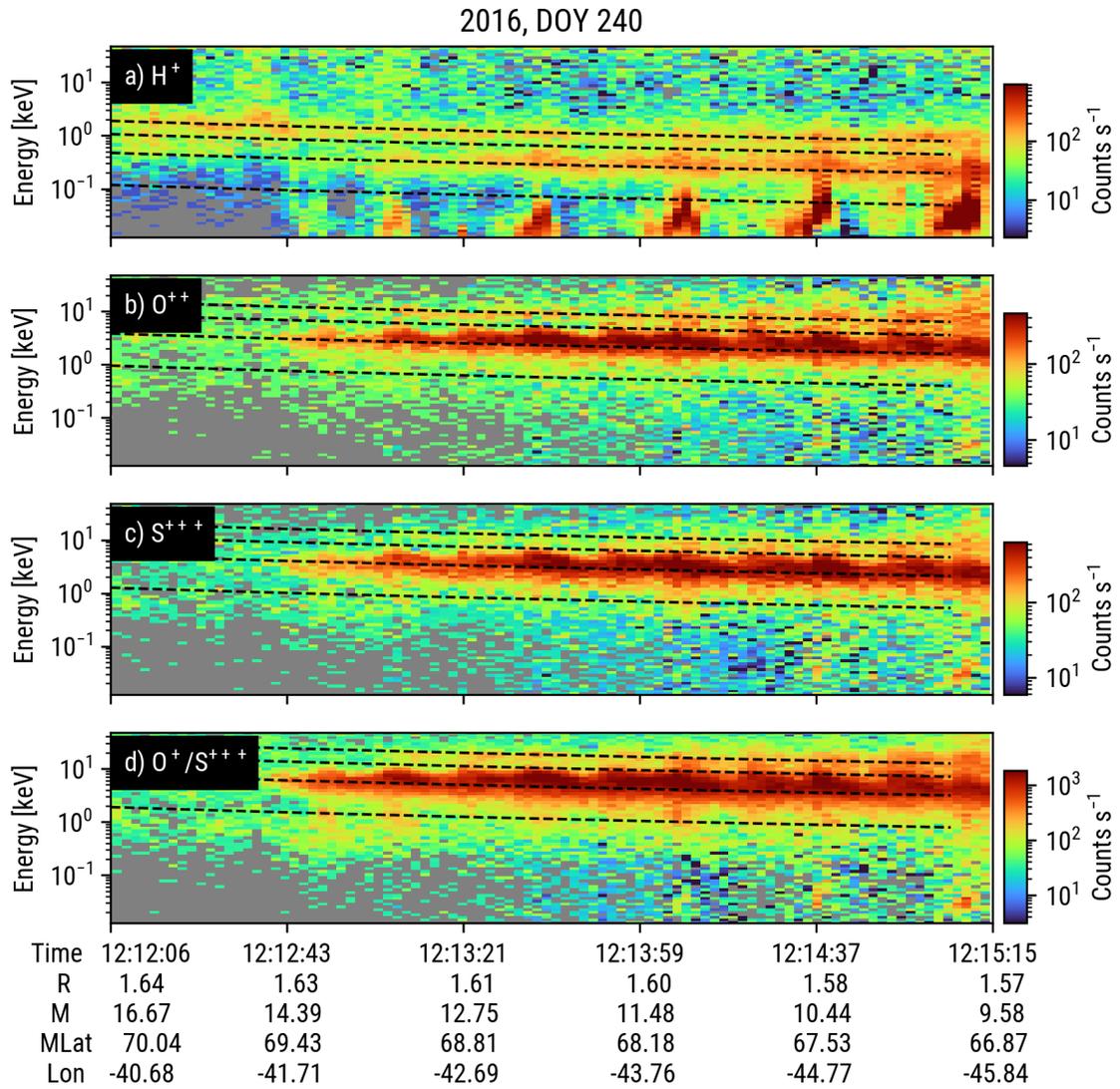

Figure S 2

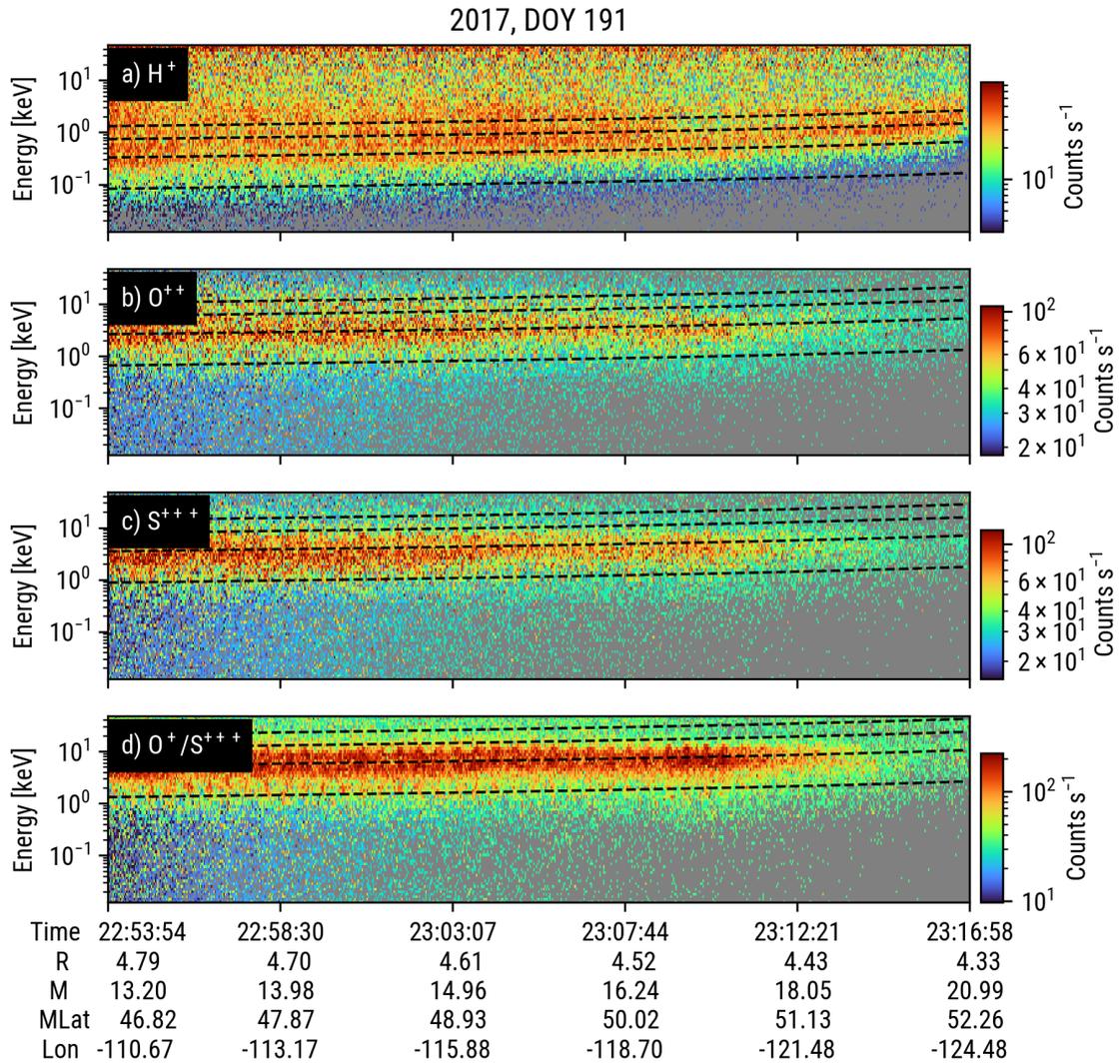

Figure S 3

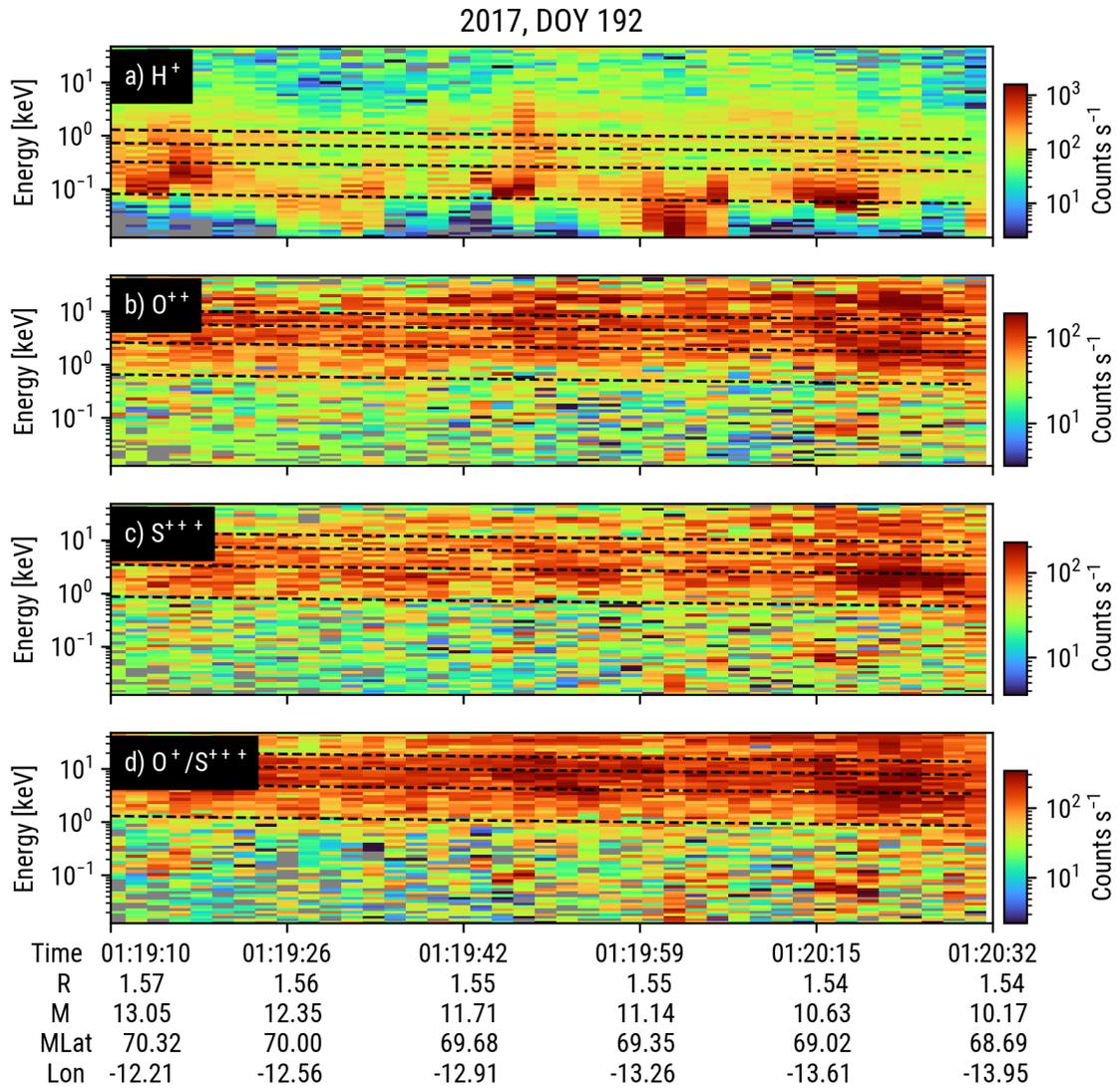

Figure S 4

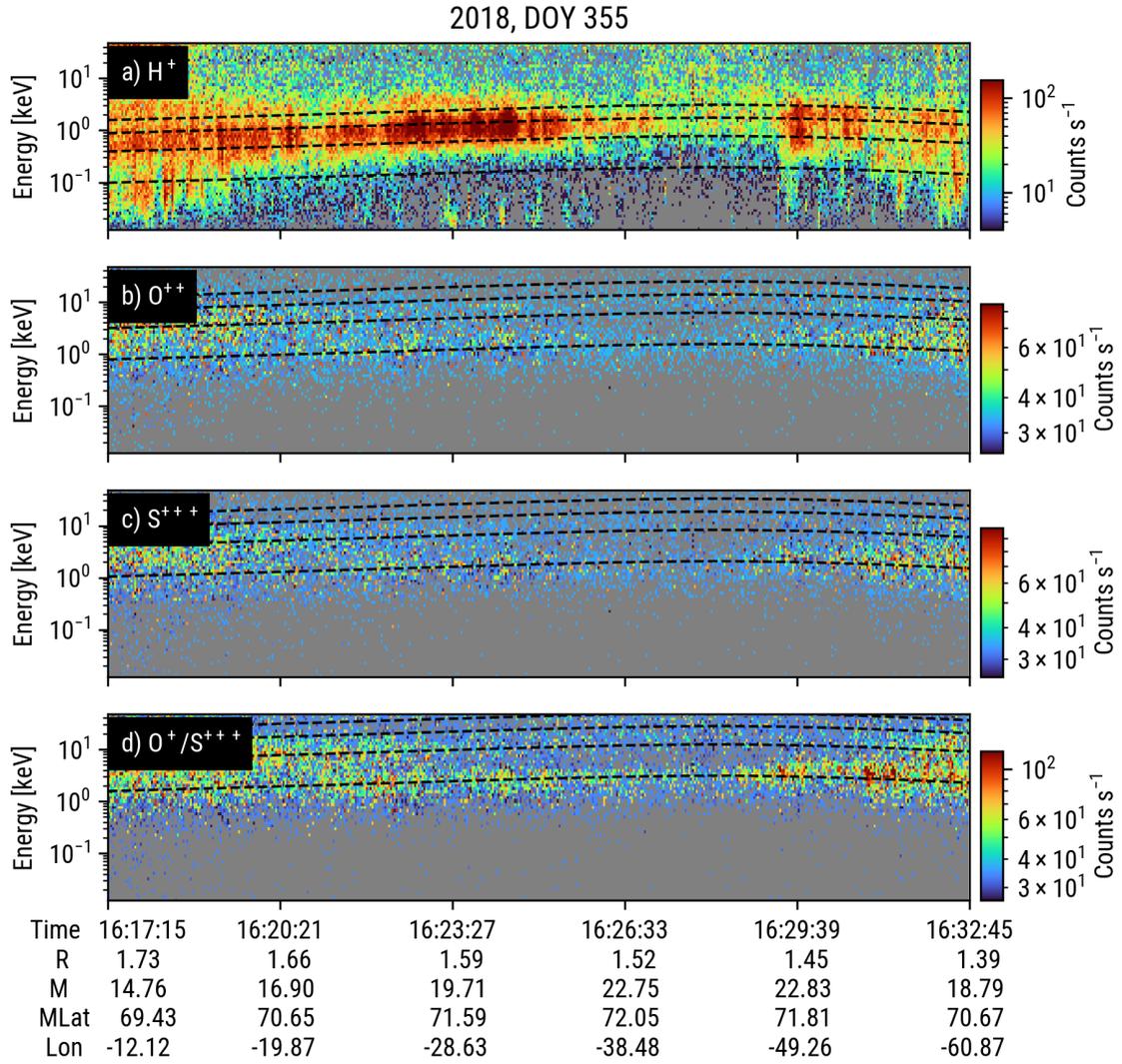

Figure S 5

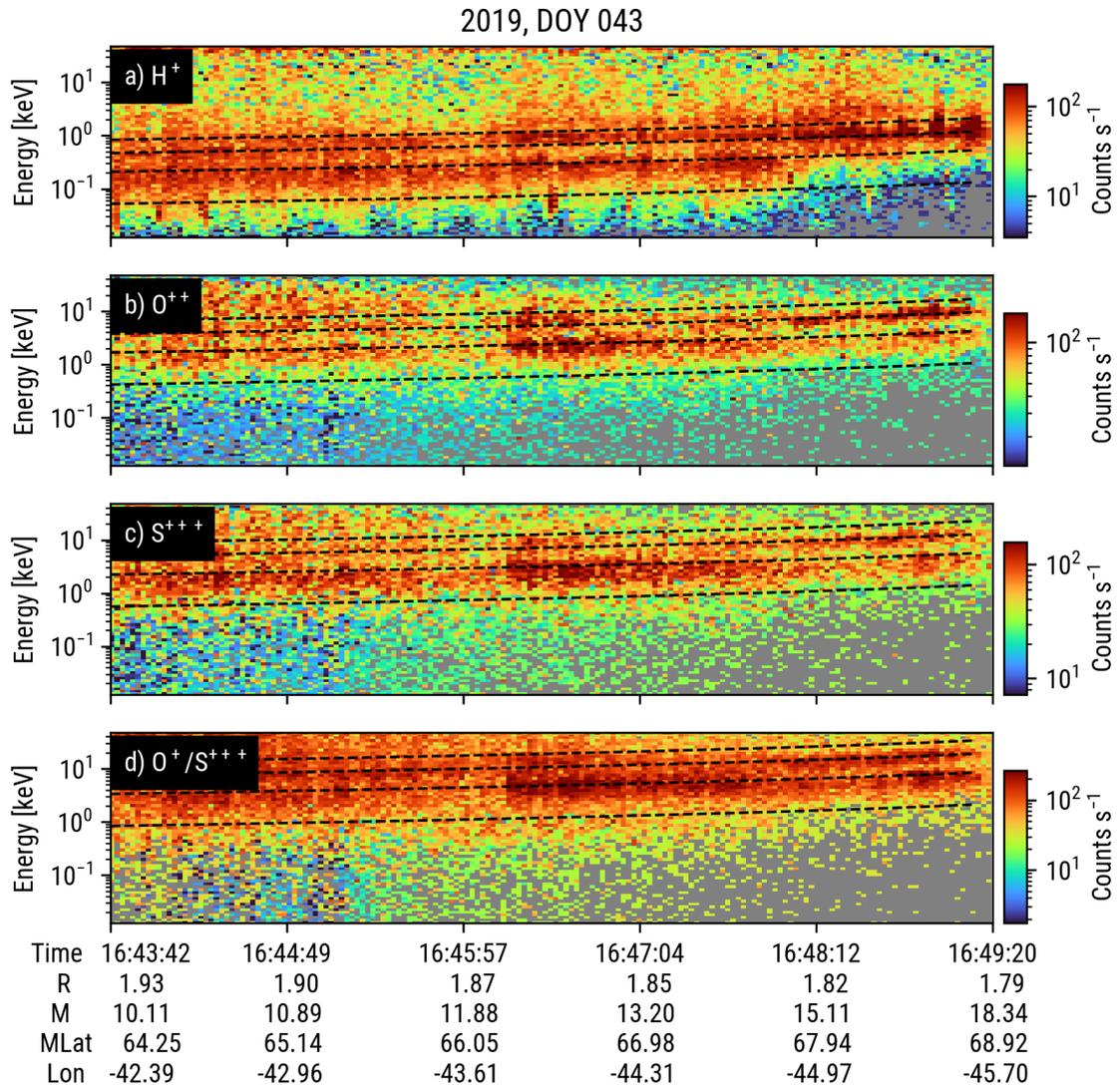

Figure S 6

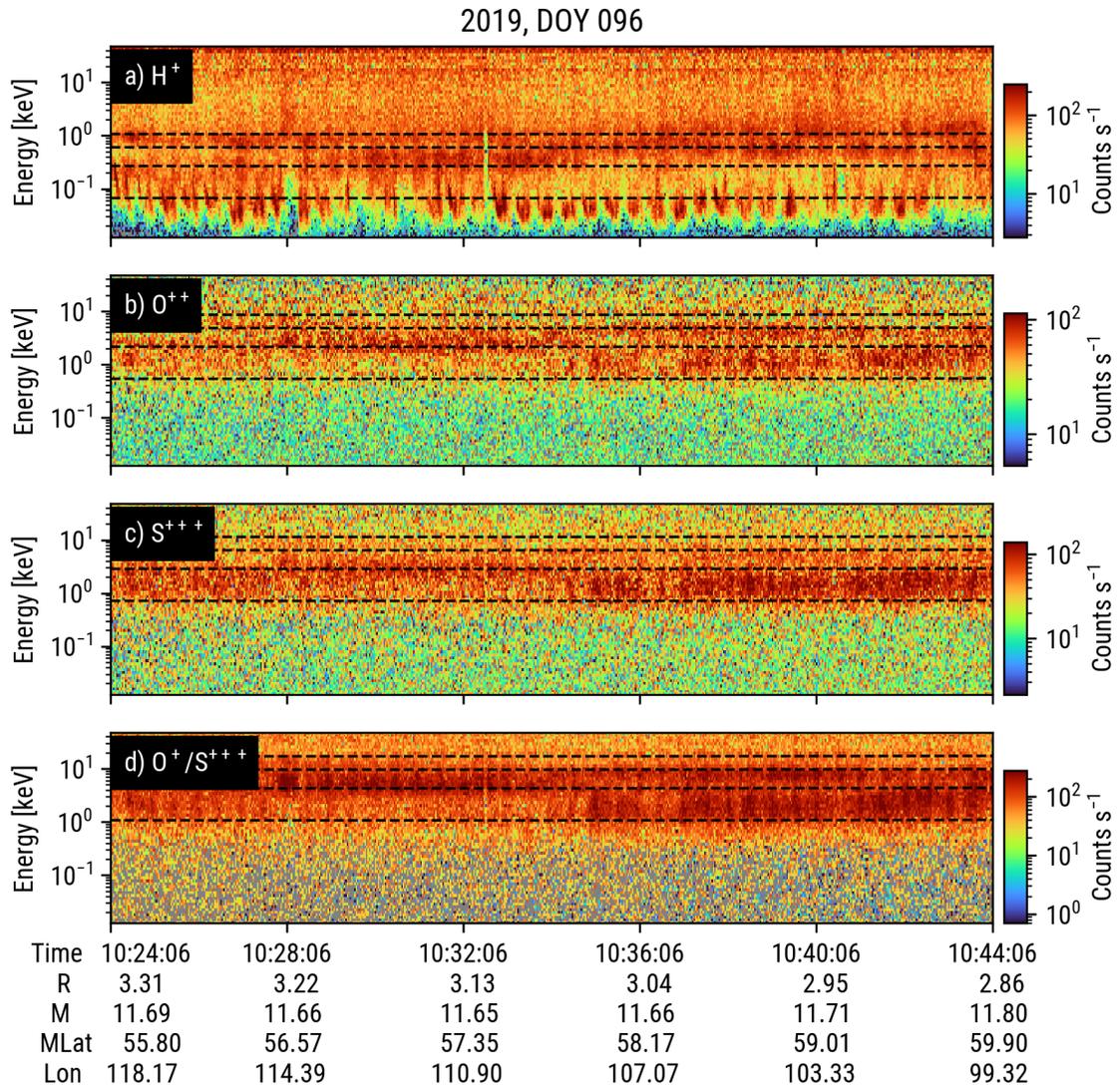

2019, DOY 096

Figure S 7

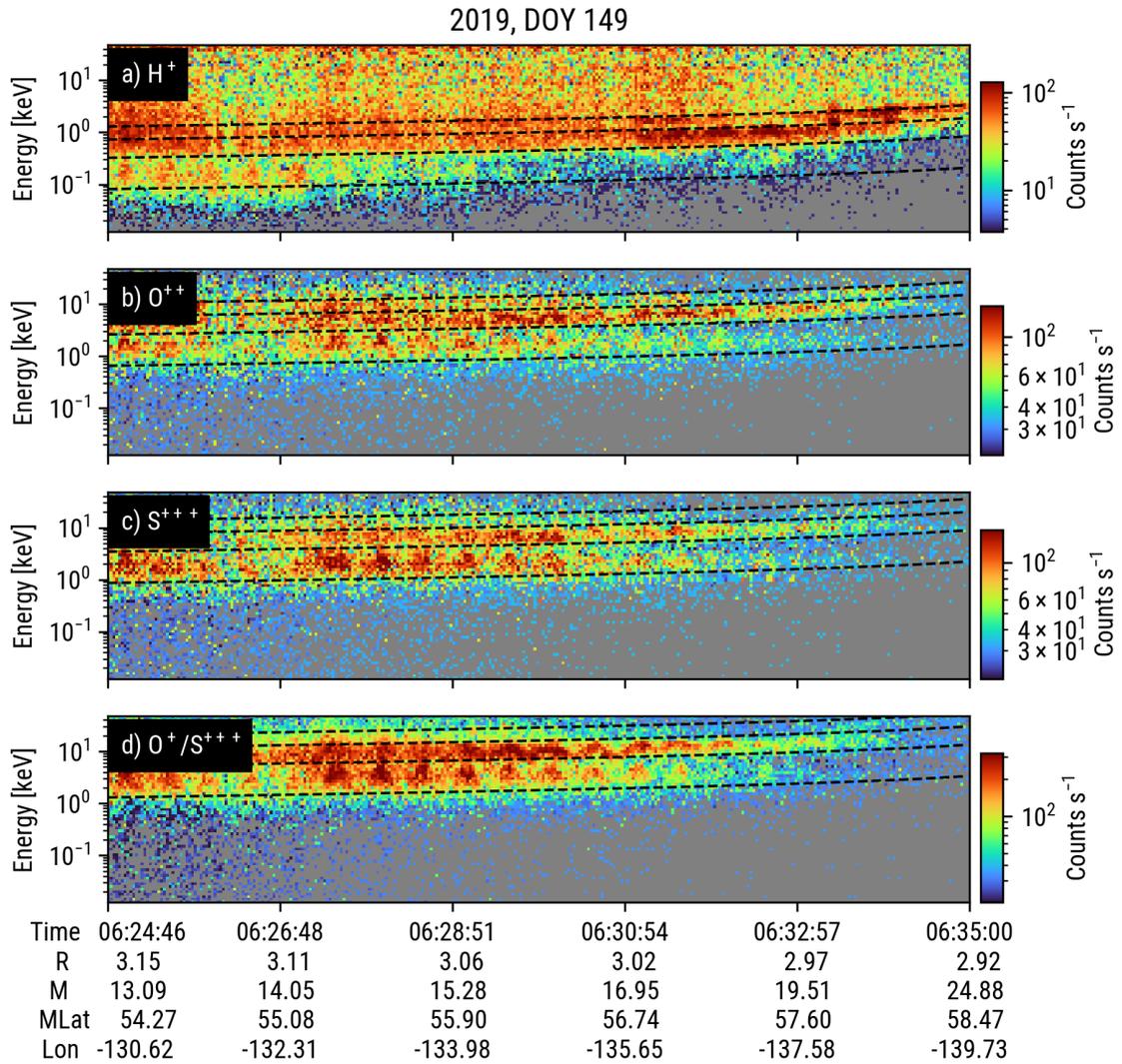

Figure S 8

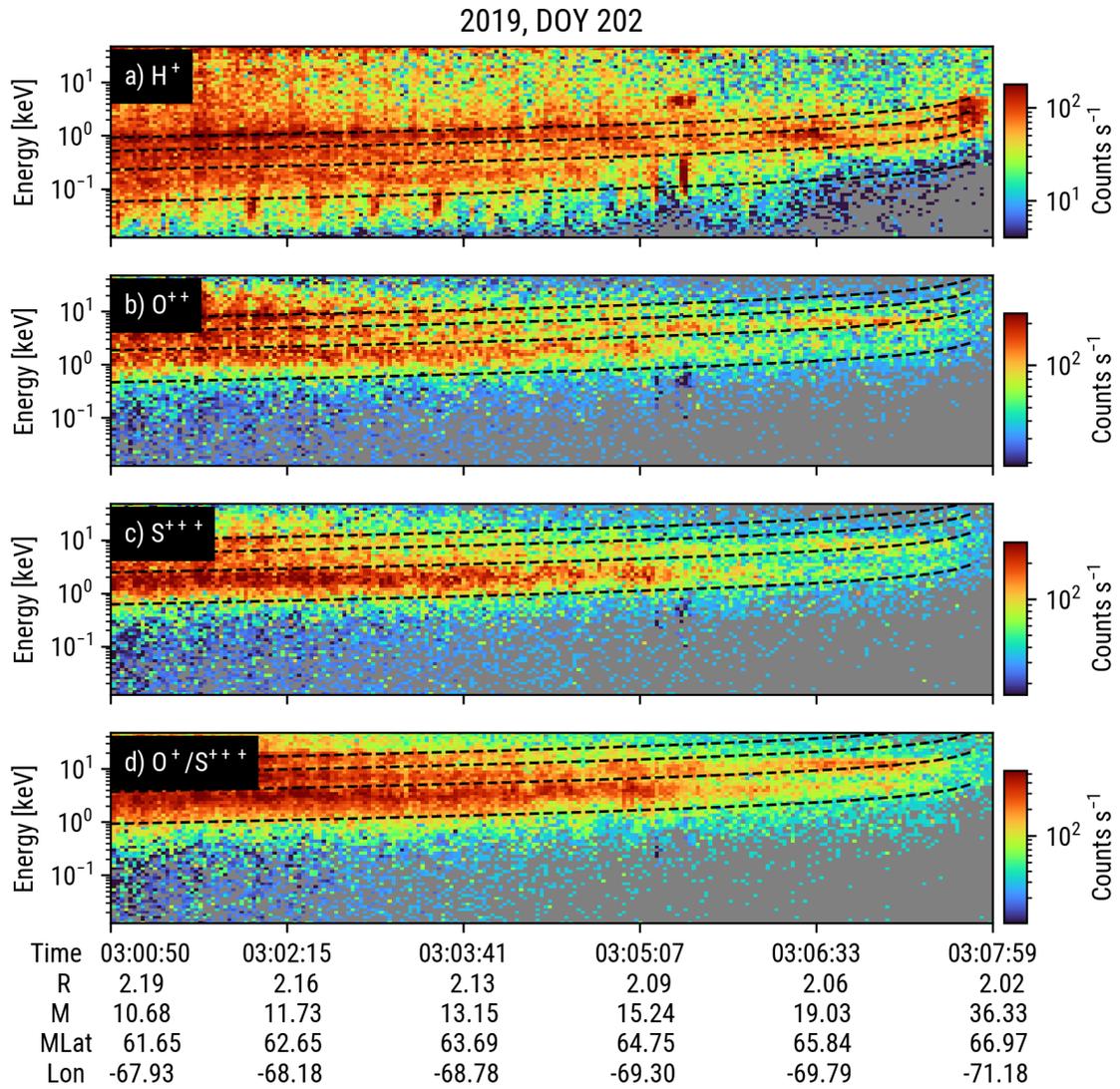

Figure S 9

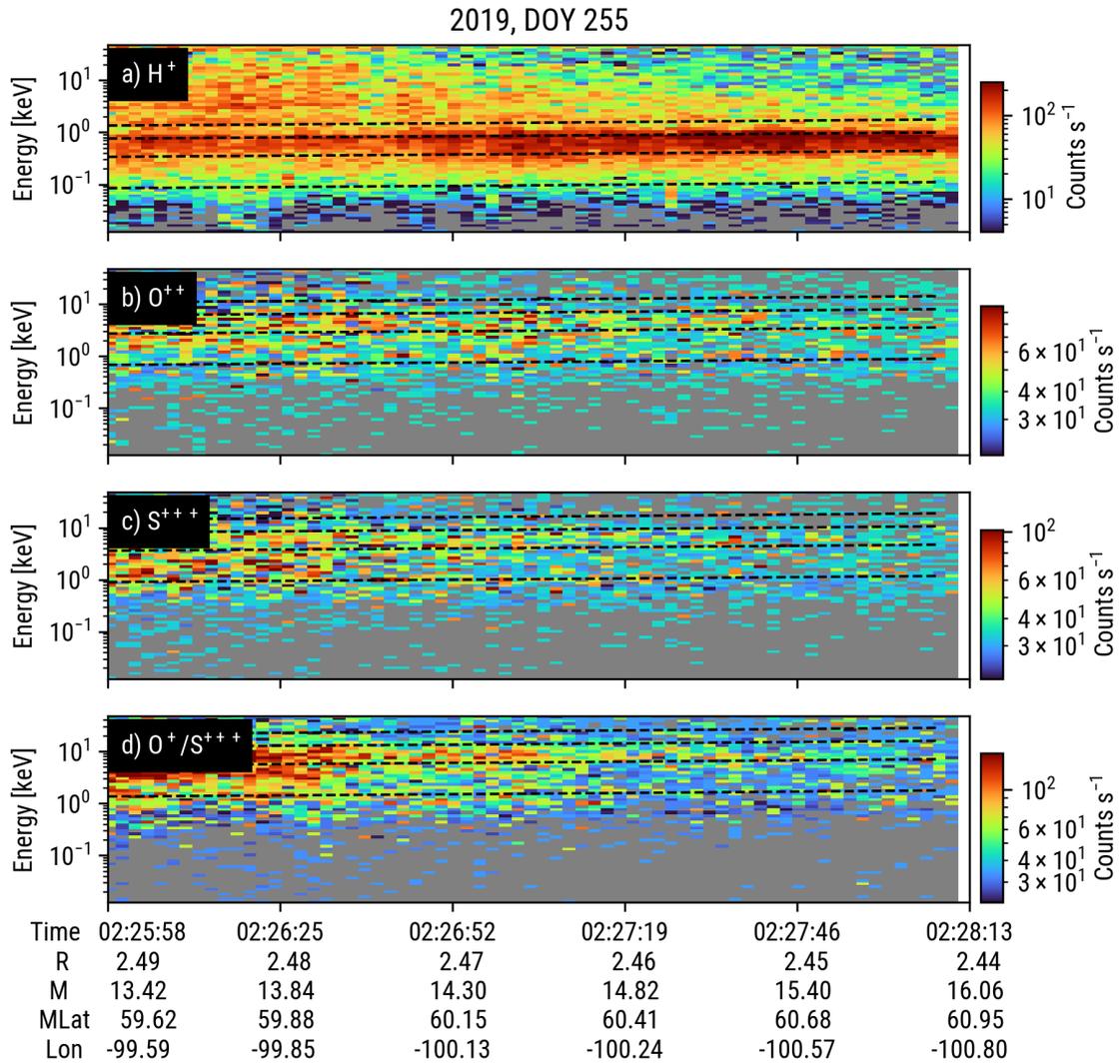

Figure S 10

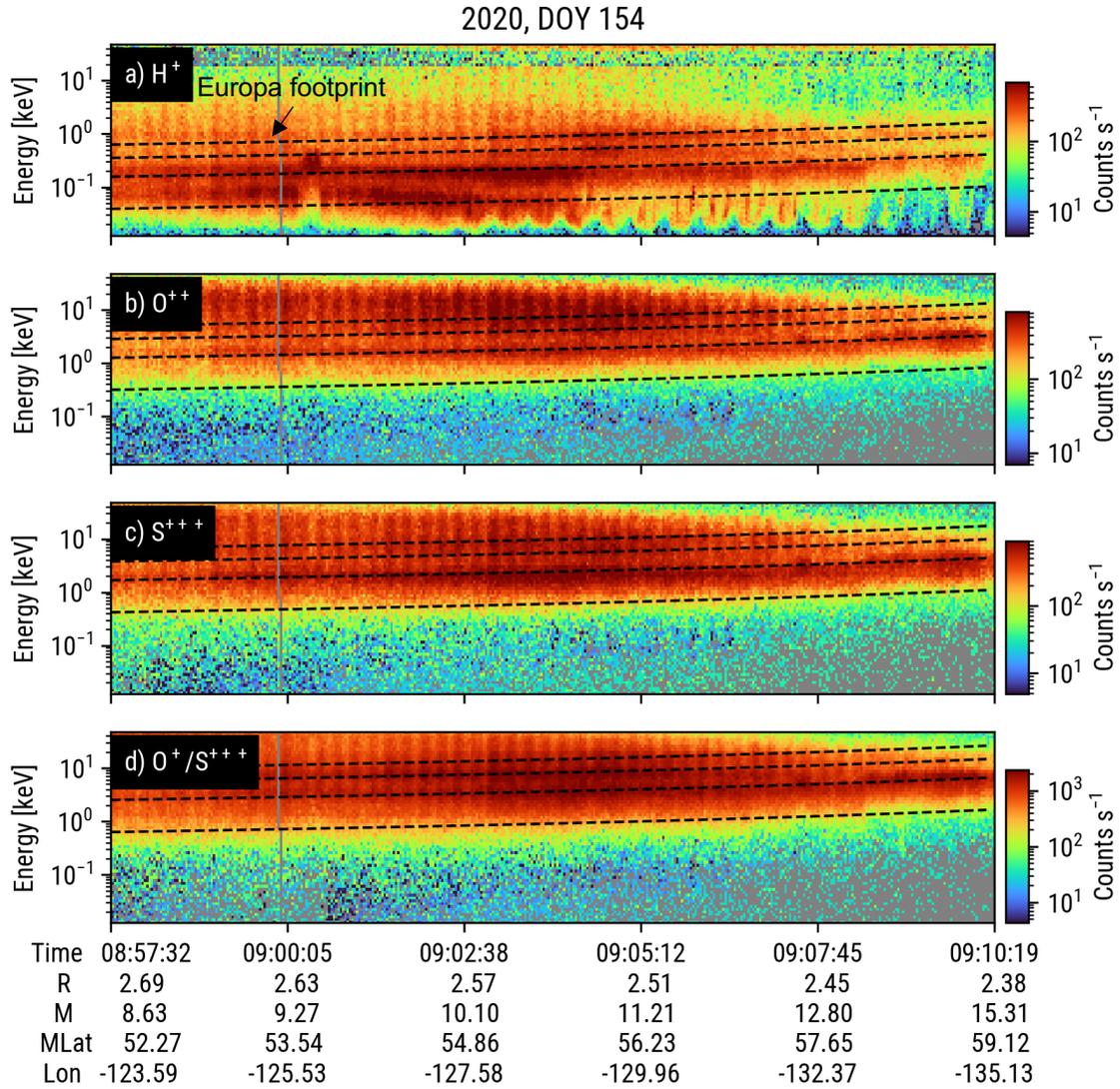

Figure S 11. Note: The Europa footprint crossing interval discussed in Sarkango et al. (2024) can also been seen here between 09:00:15 and 09:00:35 in panel (a) when the proton energy jumps to ~0.5 keV.

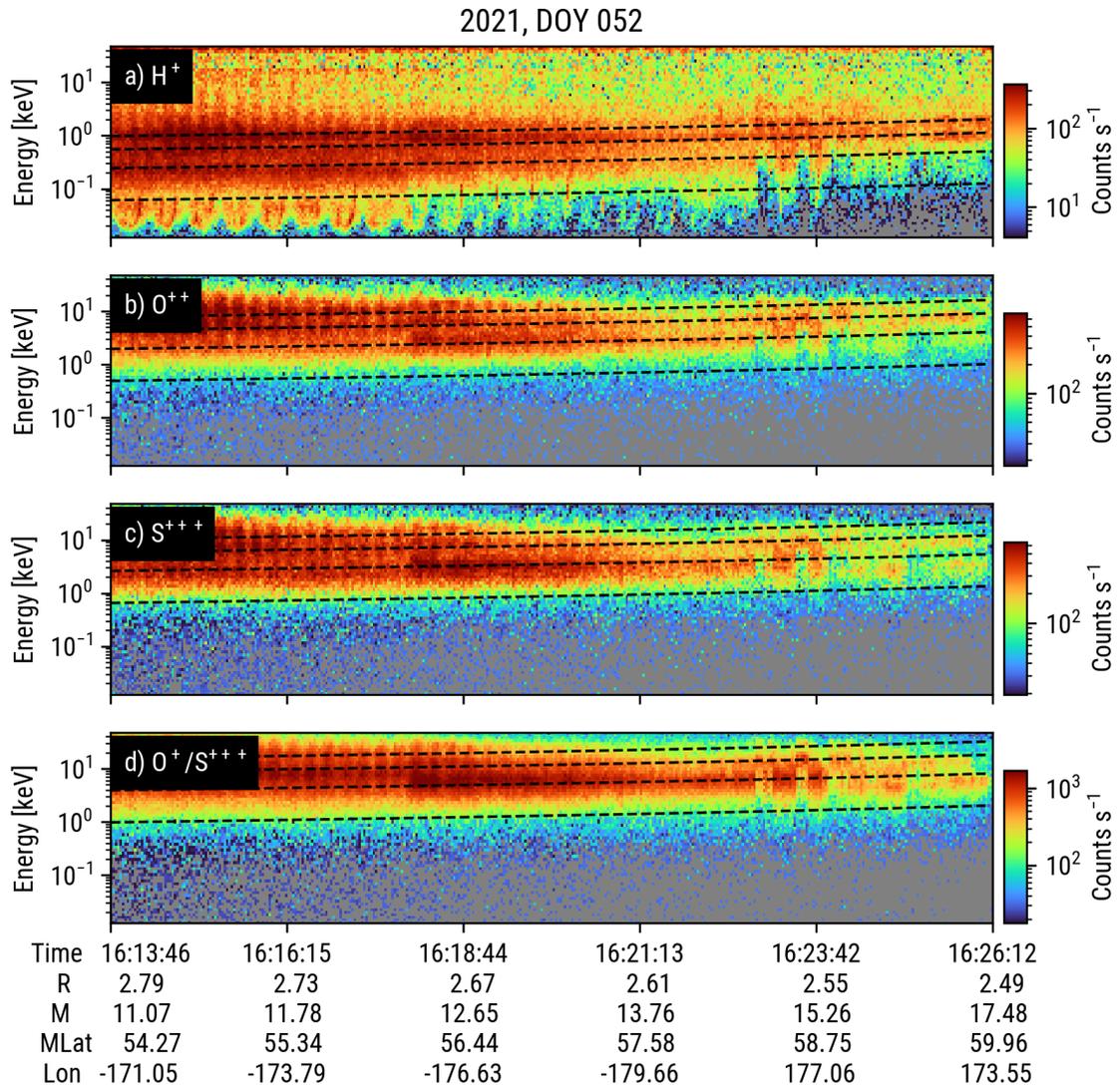

2021, DOY 052

| Time | 16:13:46 | 16:16:15 | 16:18:44 | 16:21:13 | 16:23:42 | 16:26:12 |
|------|----------|----------|----------|----------|----------|----------|
| R    | 2.79     | 2.73     | 2.67     | 2.61     | 2.55     | 2.49     |
| M    | 11.07    | 11.78    | 12.65    | 13.76    | 15.26    | 17.48    |
| MLat | 54.27    | 55.34    | 56.44    | 57.58    | 58.75    | 59.96    |
| Lon  | -171.05  | -173.79  | -176.63  | -179.66  | 177.06   | 173.55   |

Figure S 12

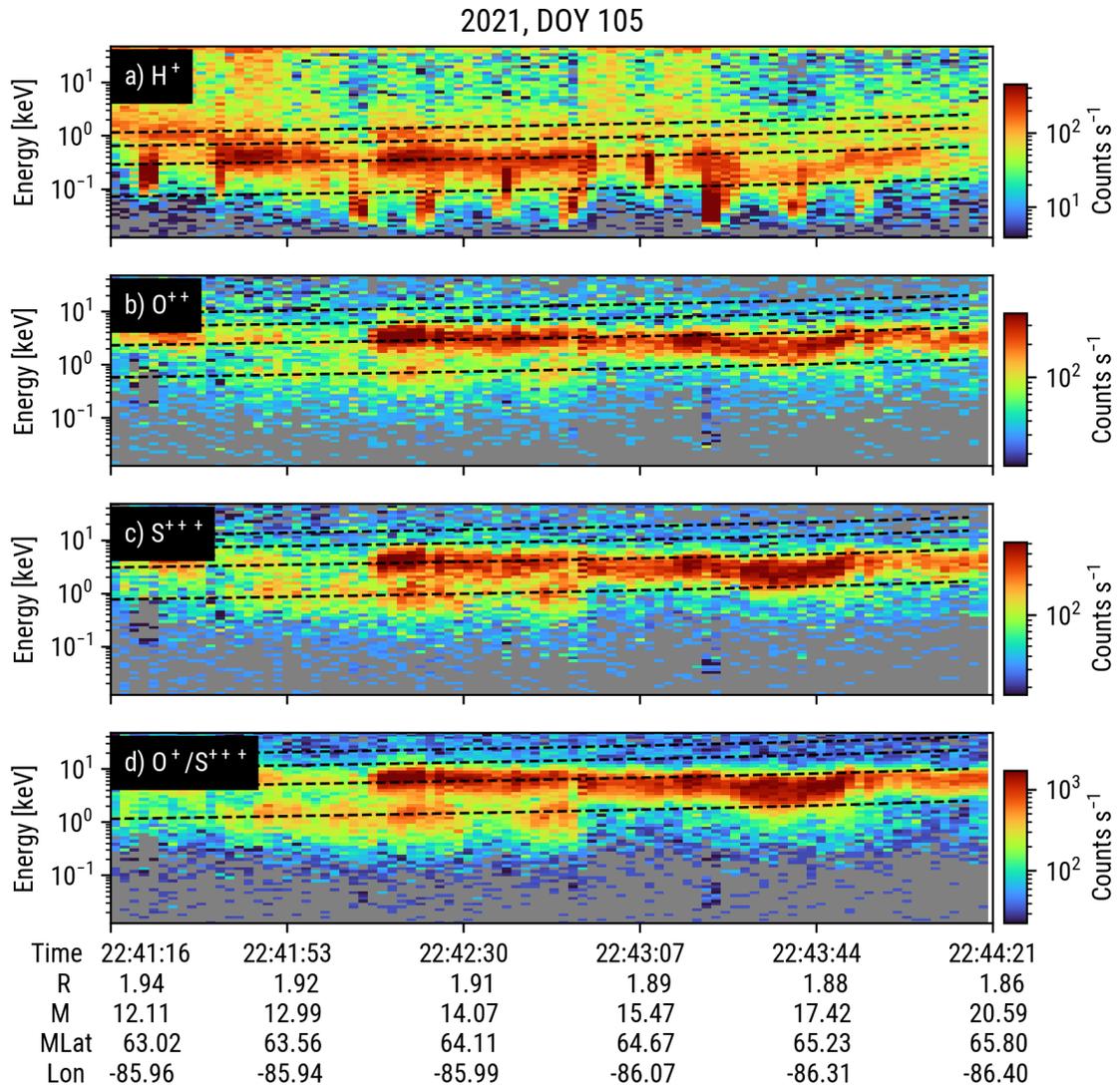

Figure S 13

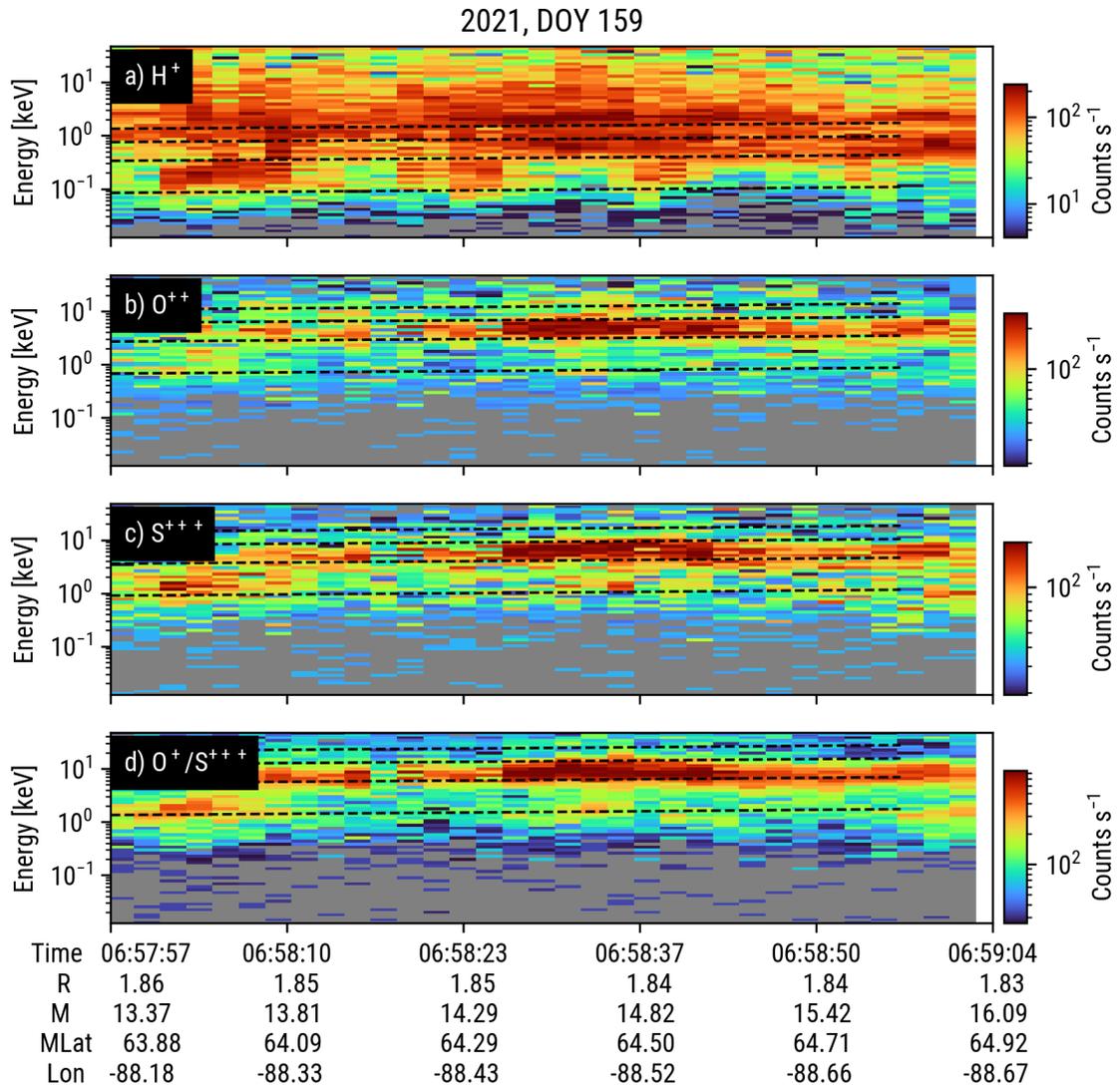

Figure S 14

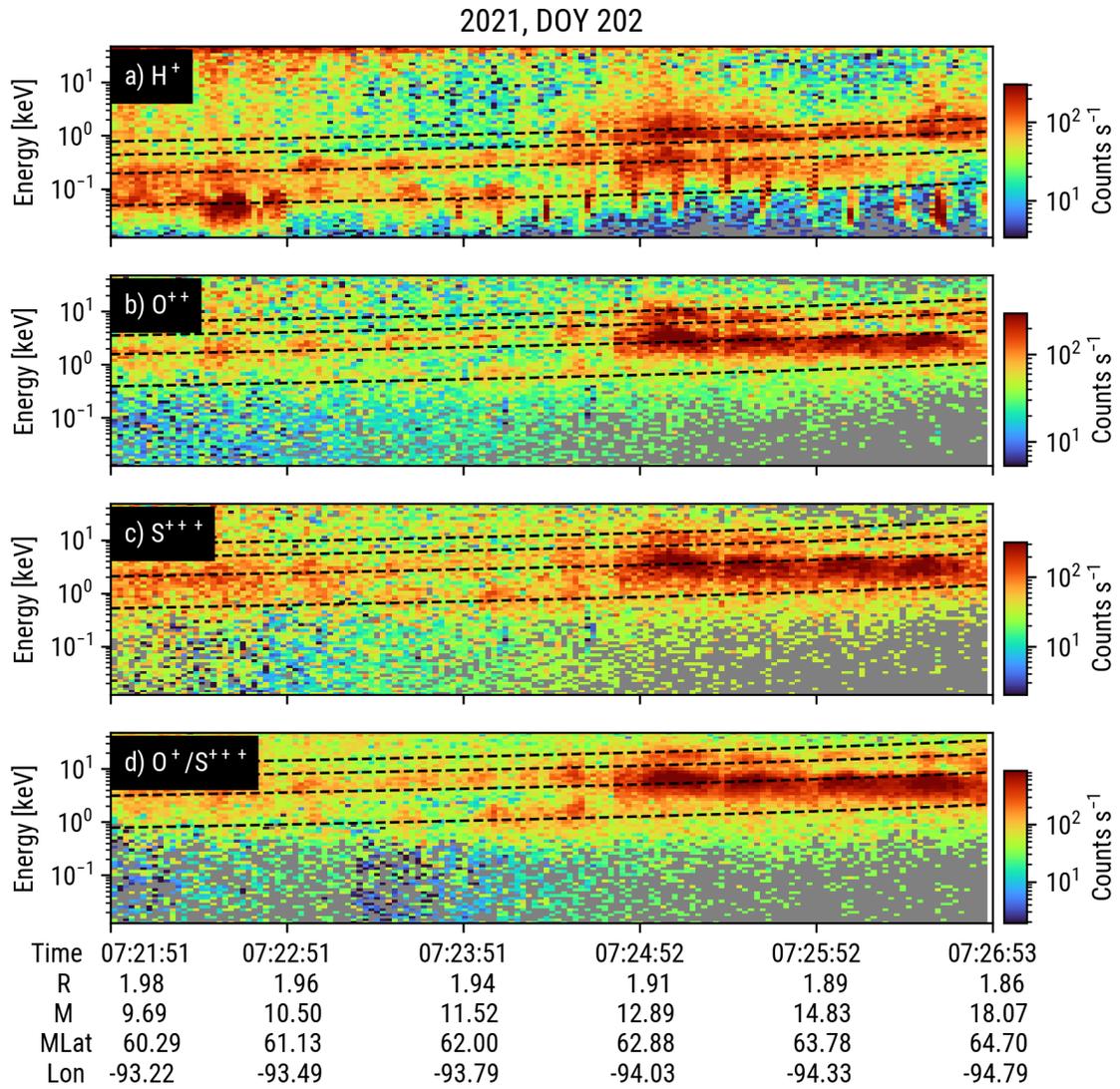

Figure S 15

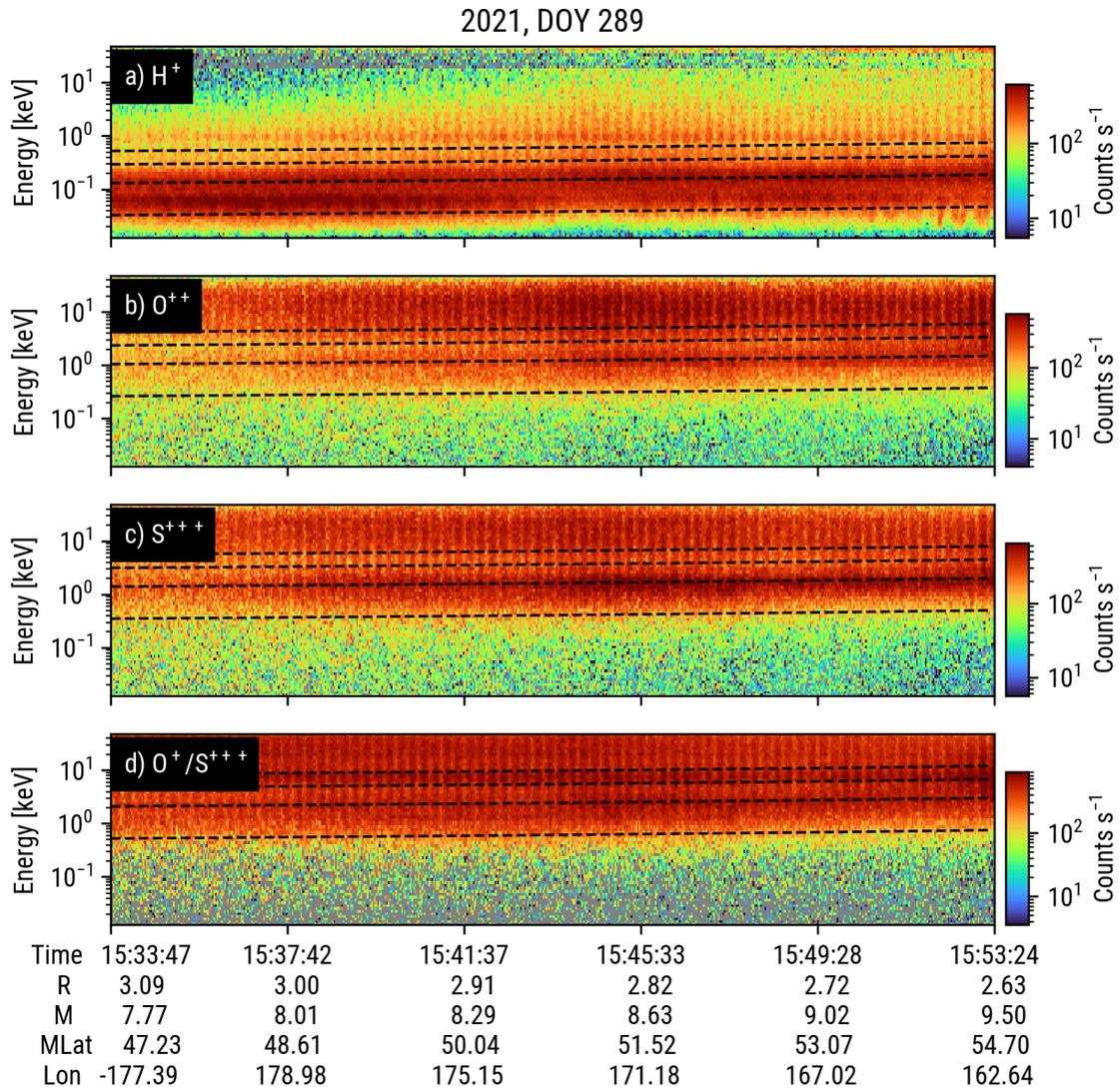

Figure S 16

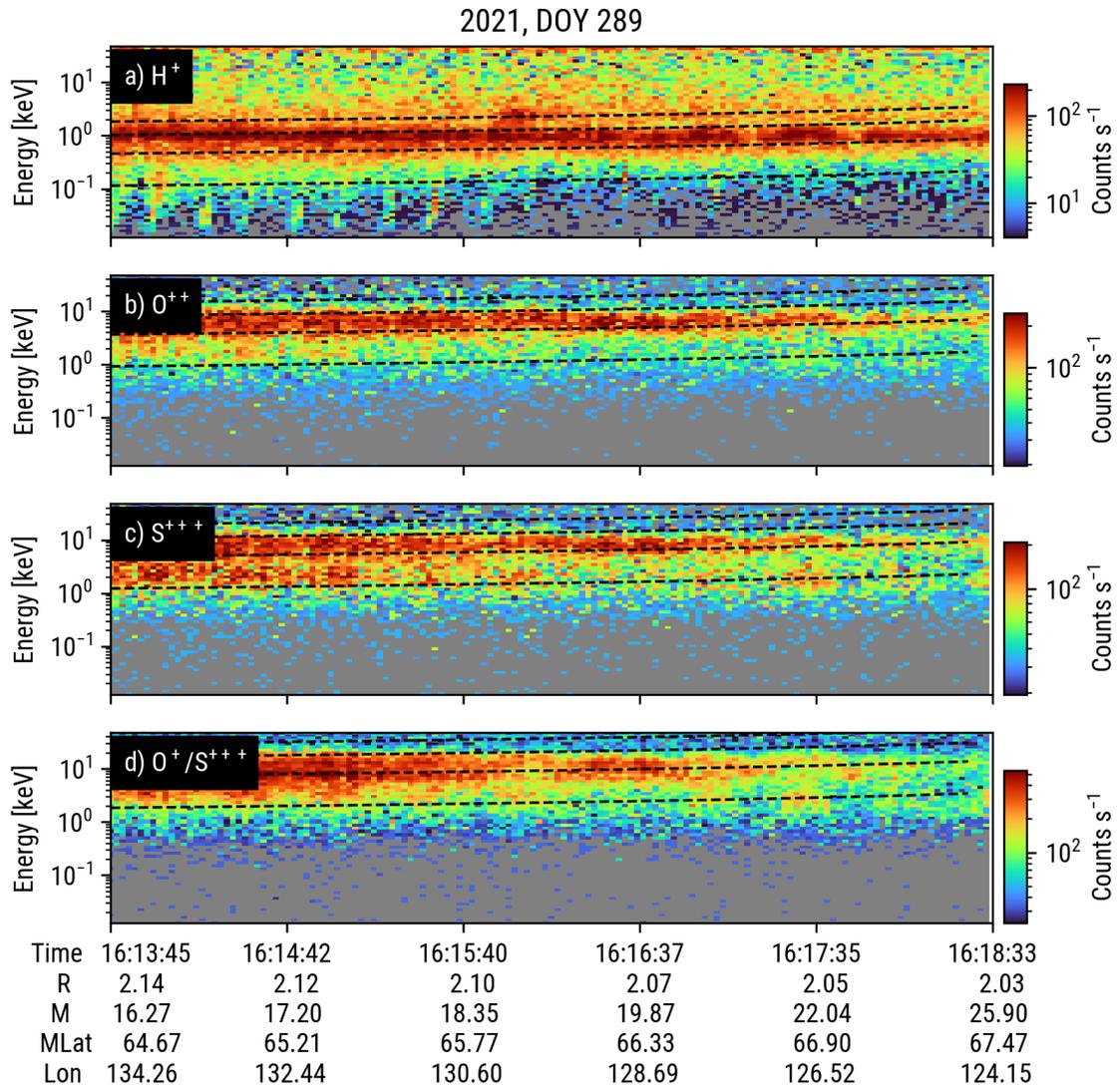

Figure S 17

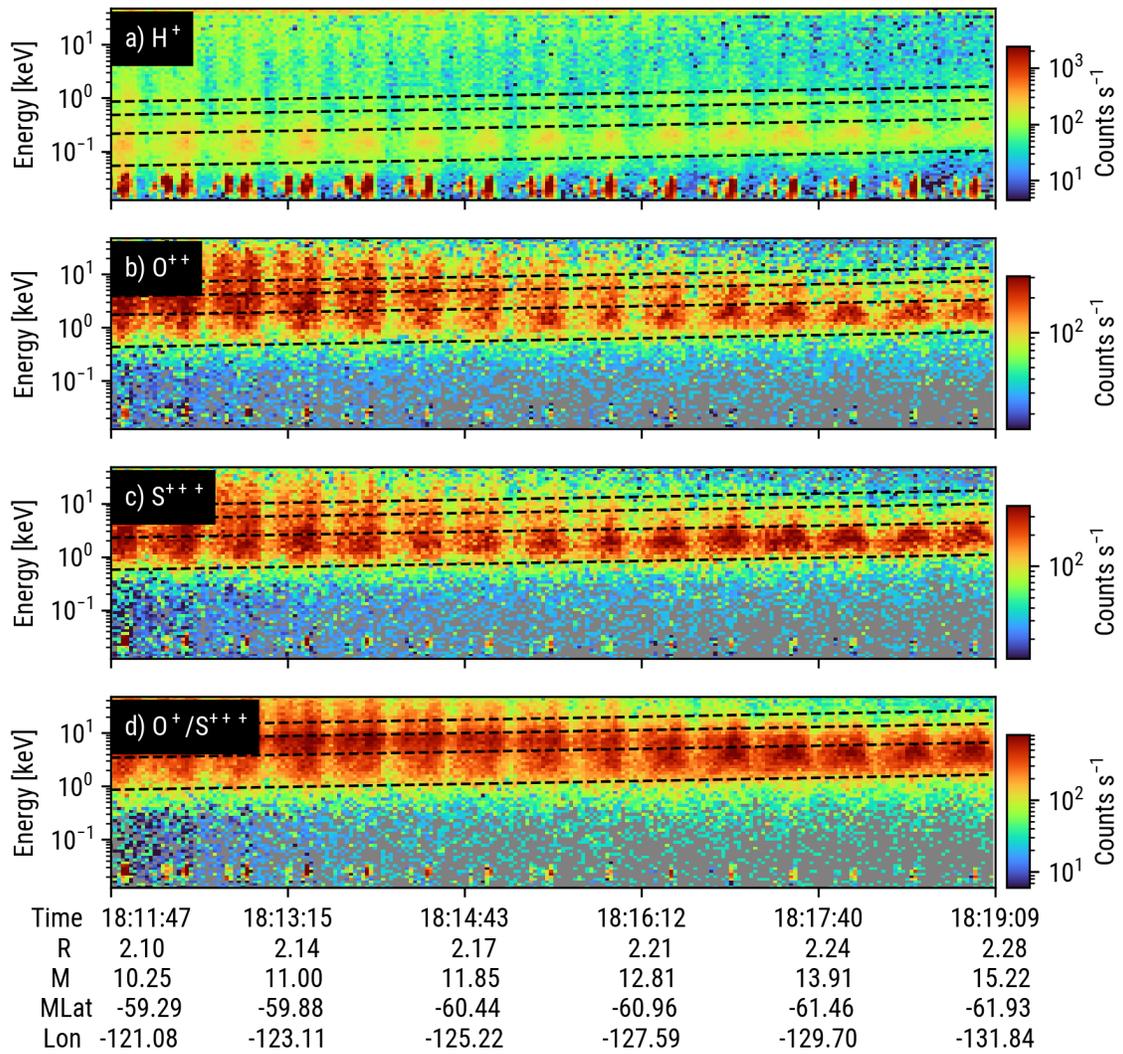

2021, DOY 289

Figure S 18

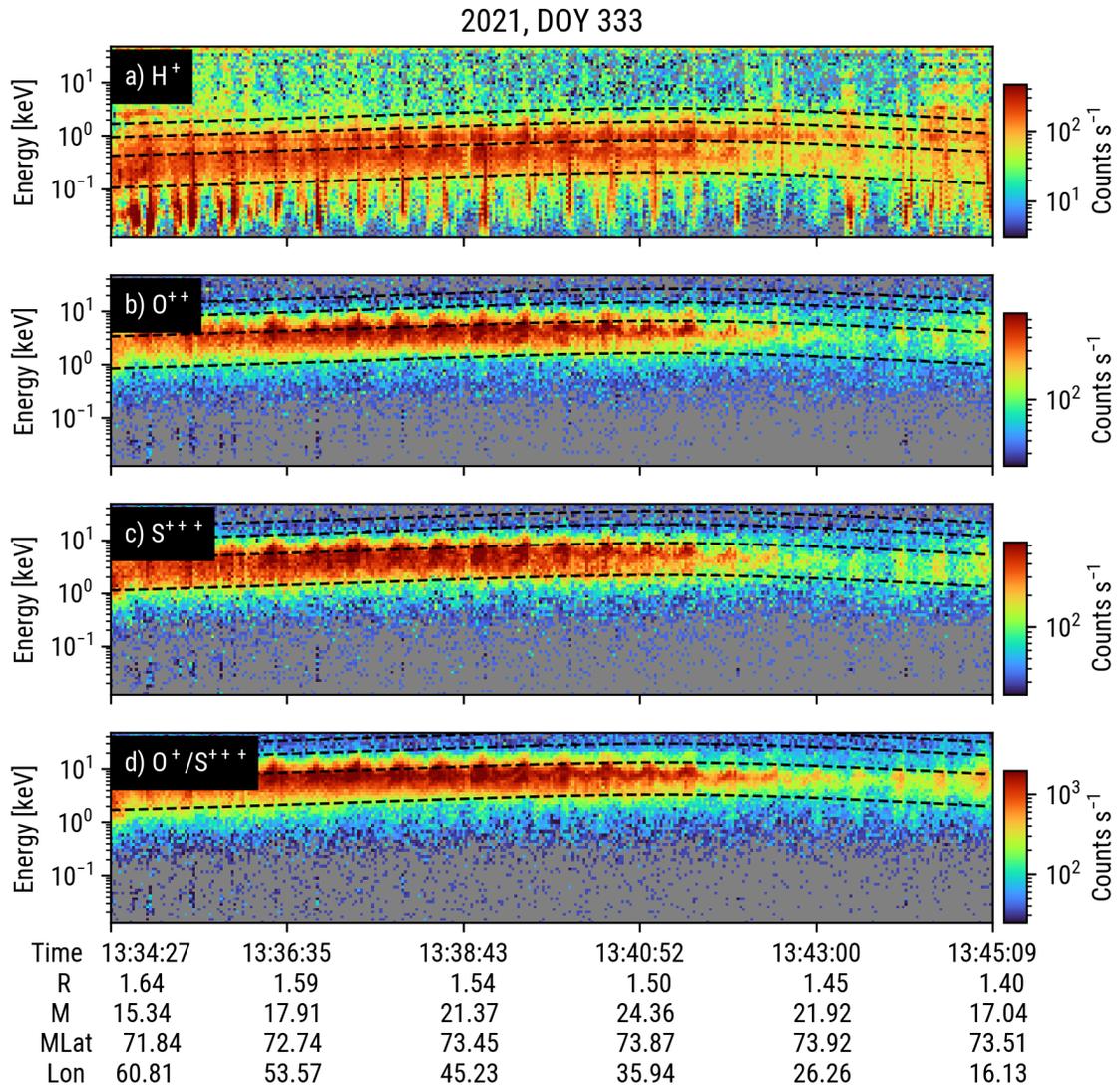

Figure S 19

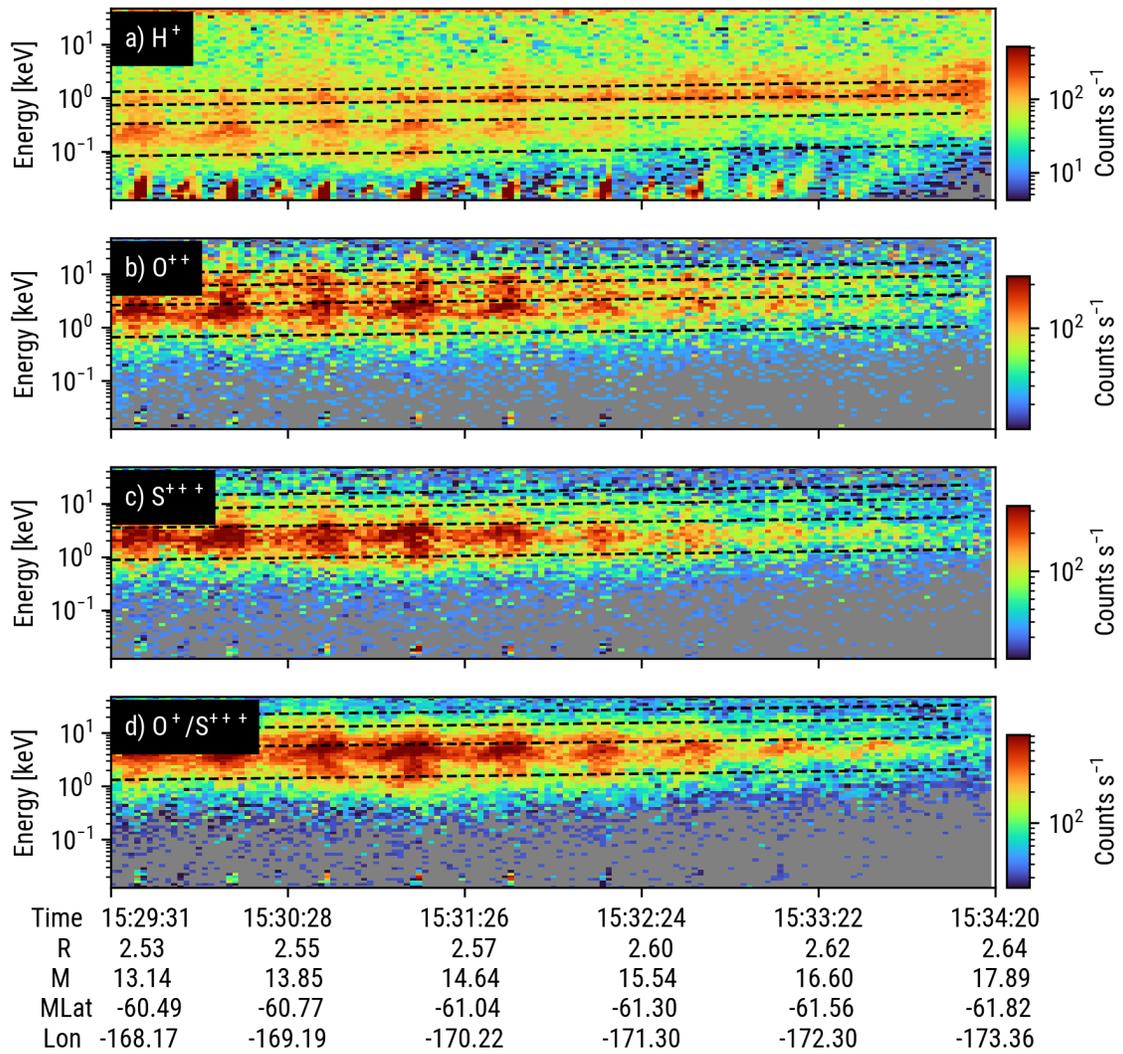

Figure S 20

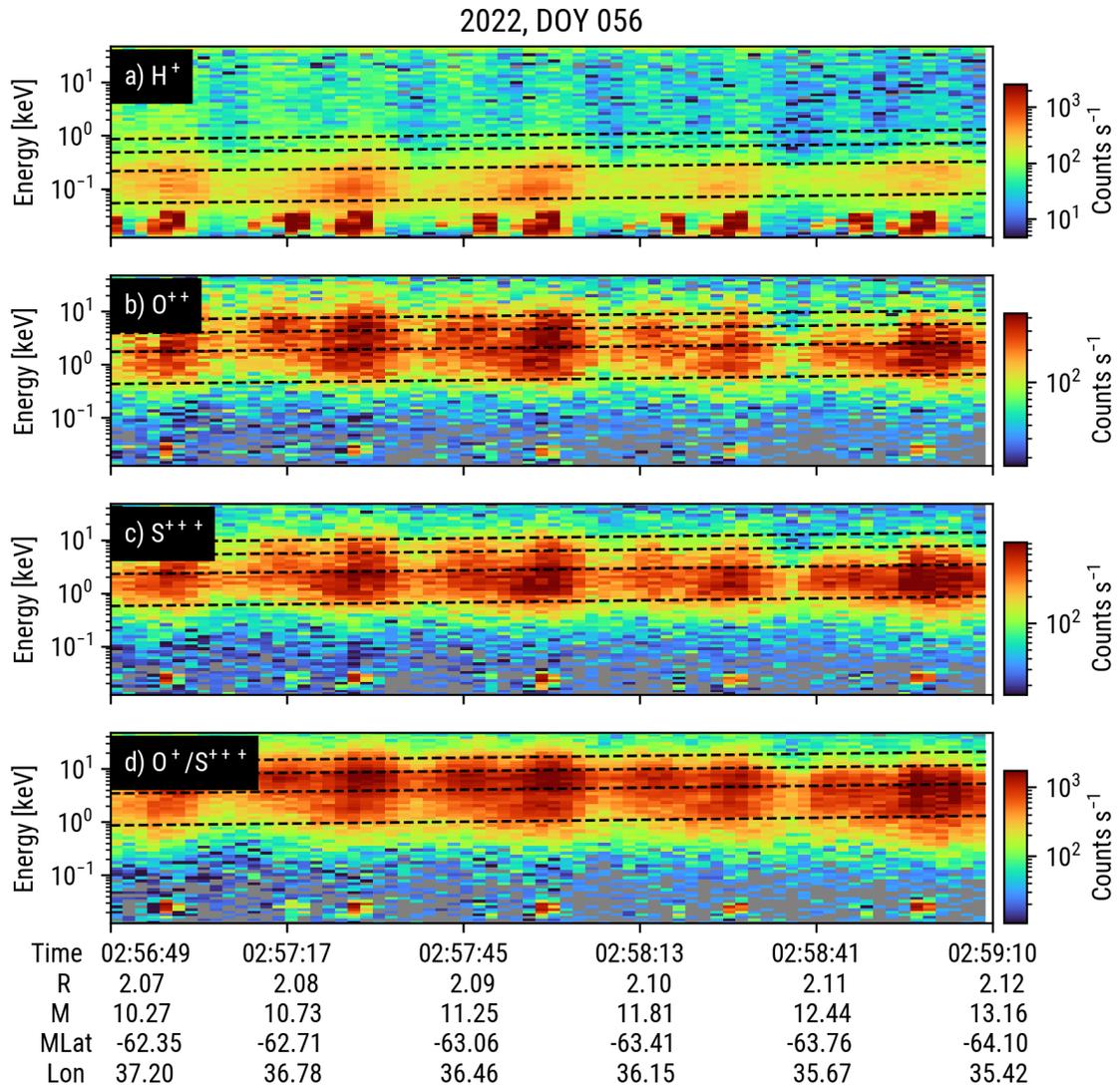

Figure S 21

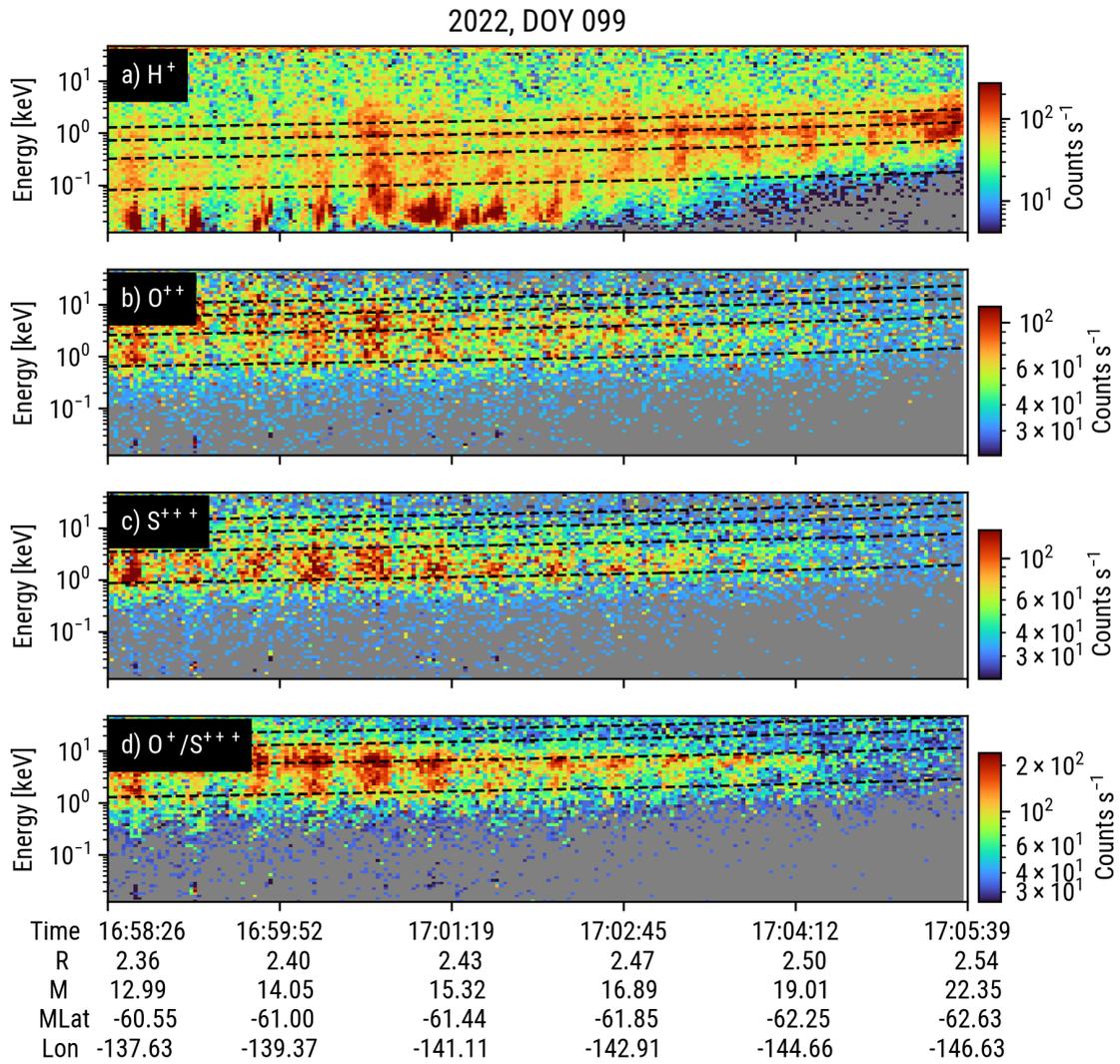

Figure S 22

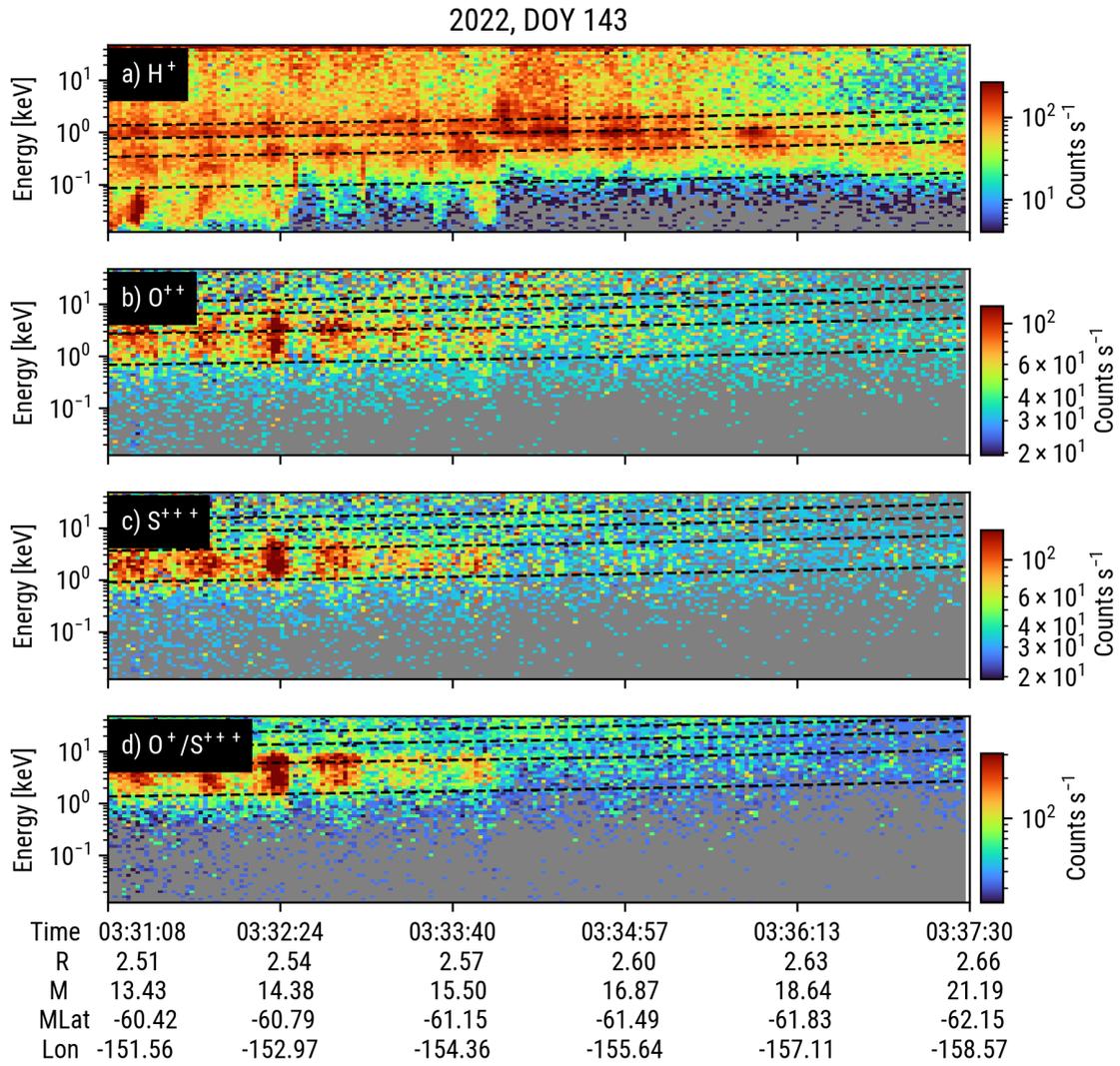

Figure S 23

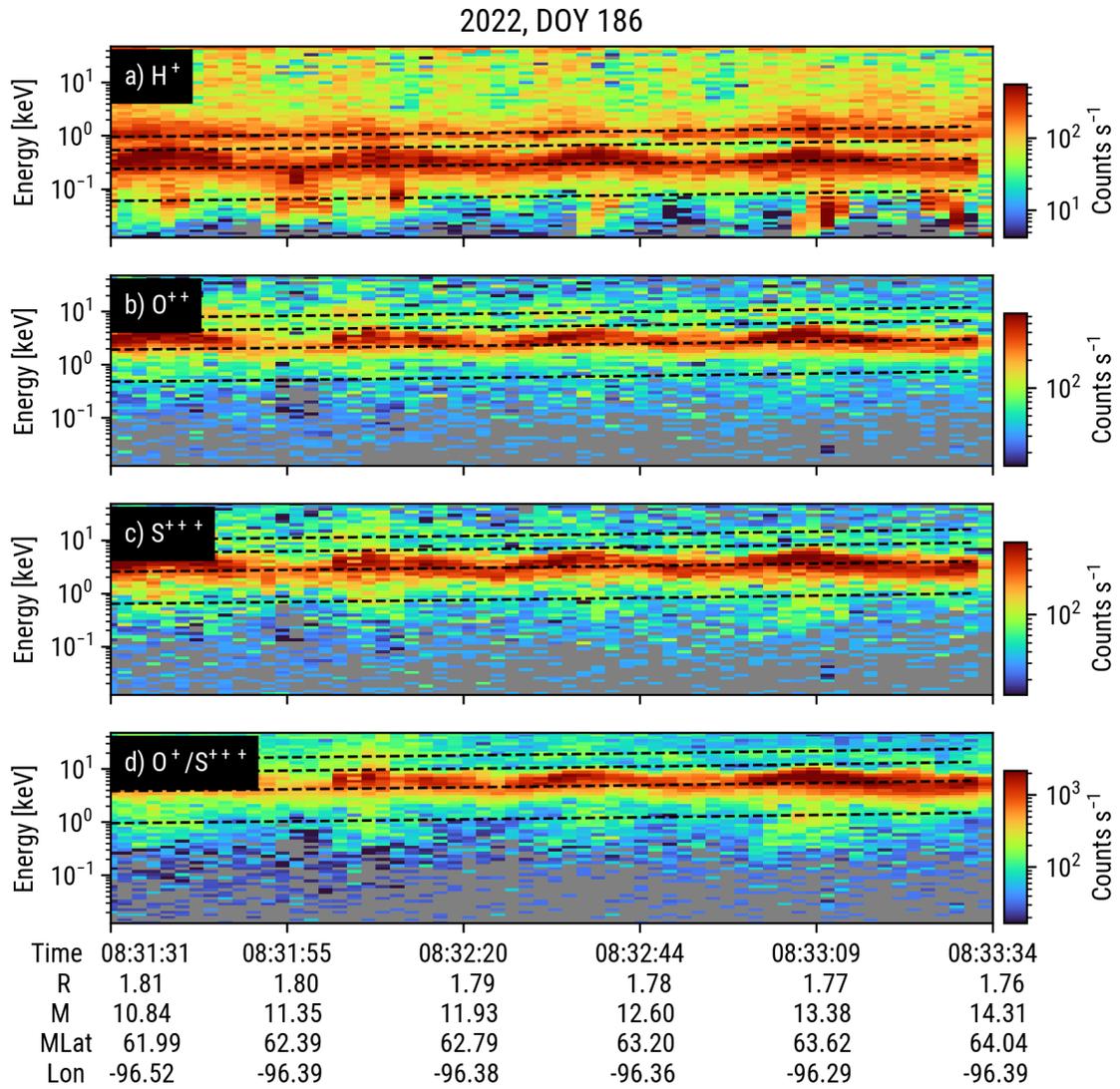

2022, DOY 186

Figure S 24

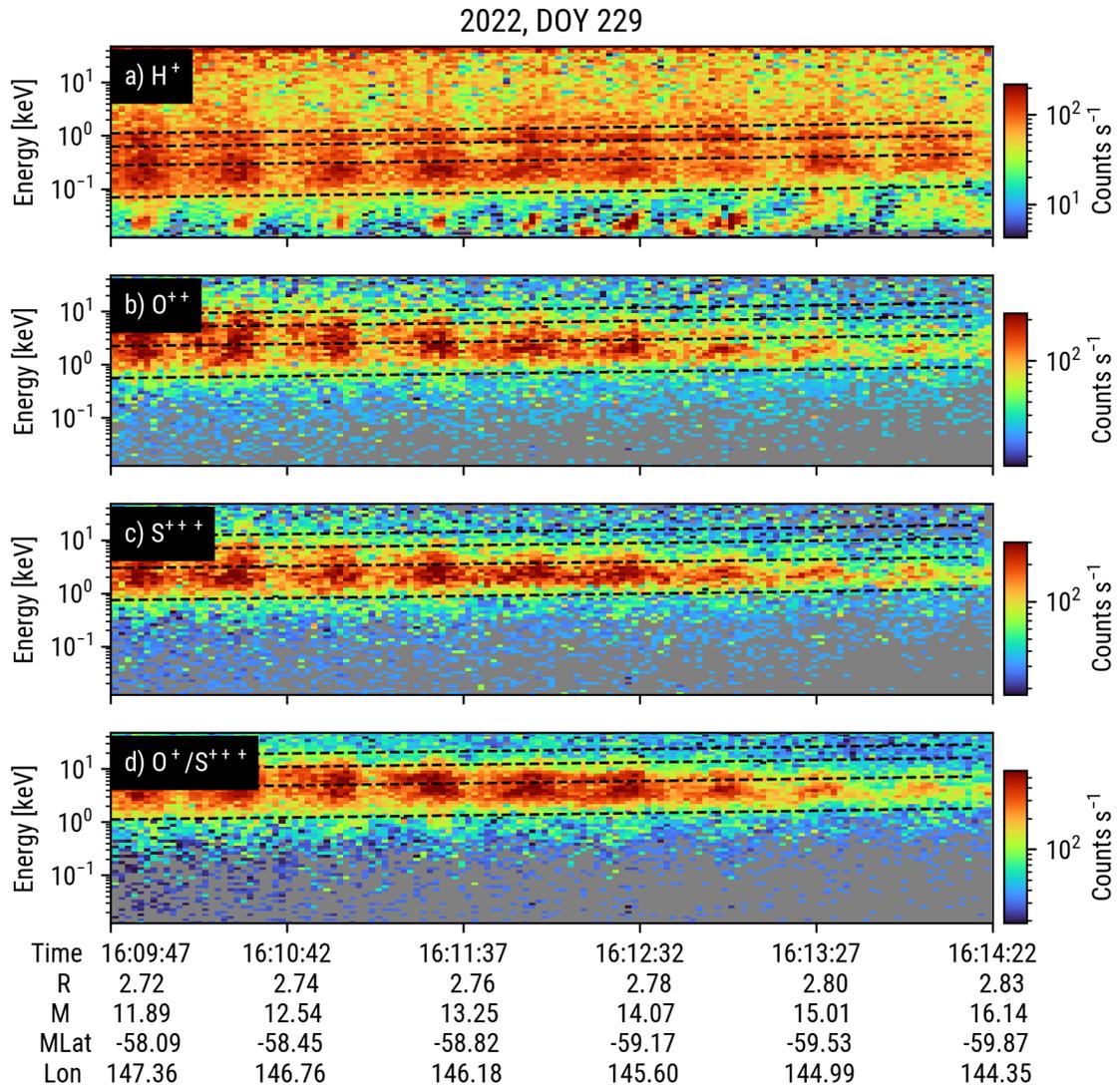

Figure S 25

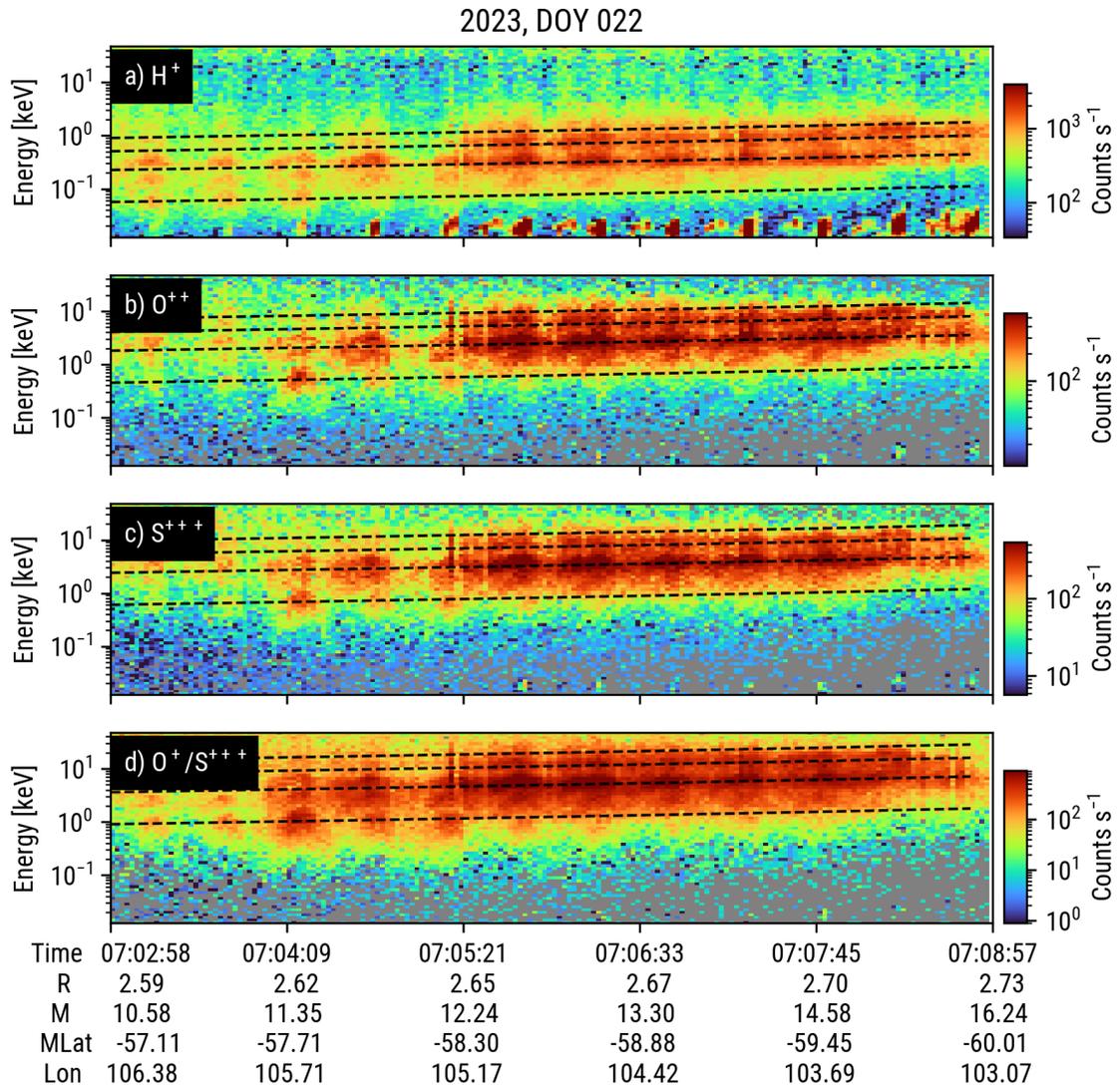

Figure S 26

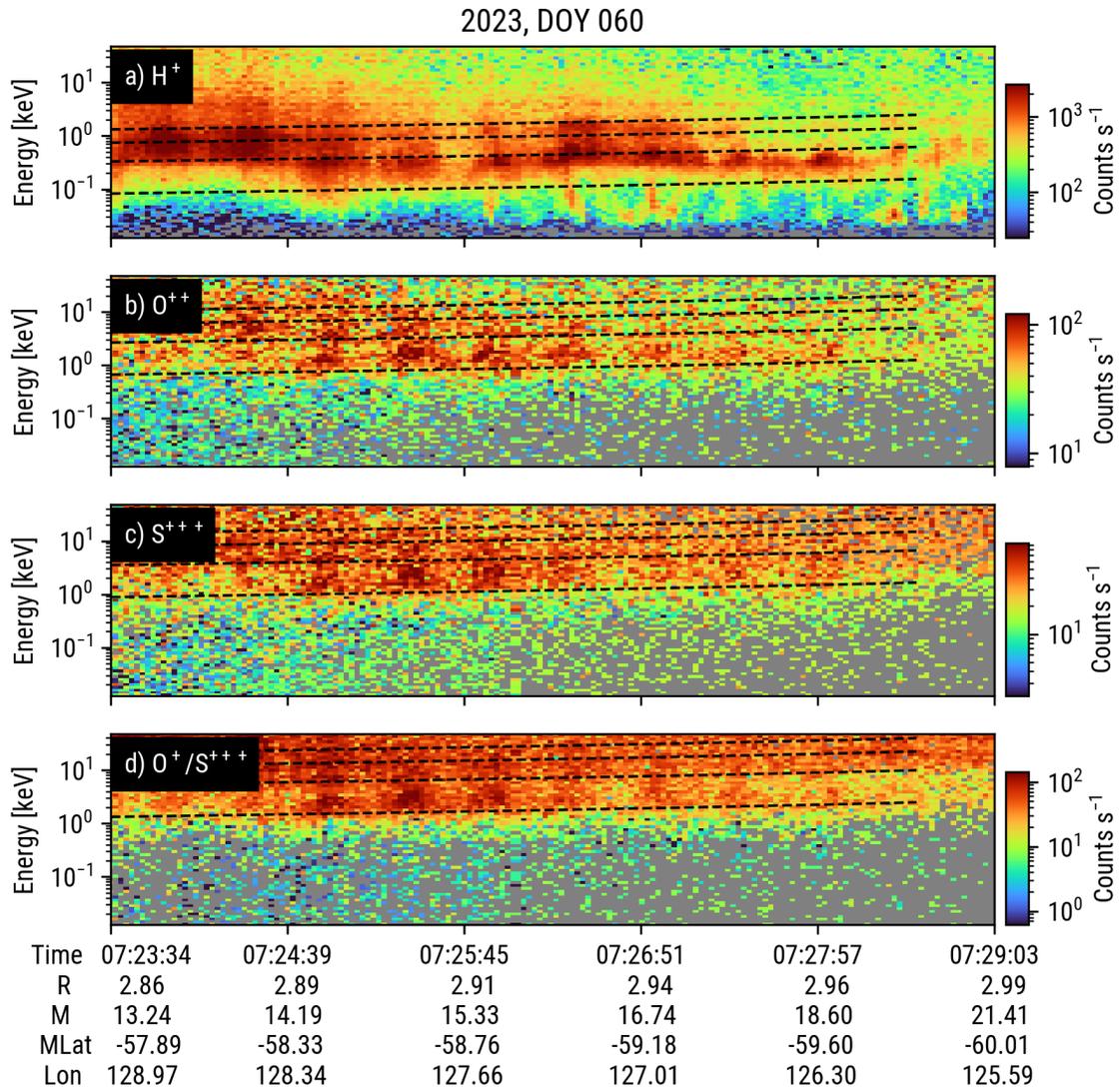

Figure S 27

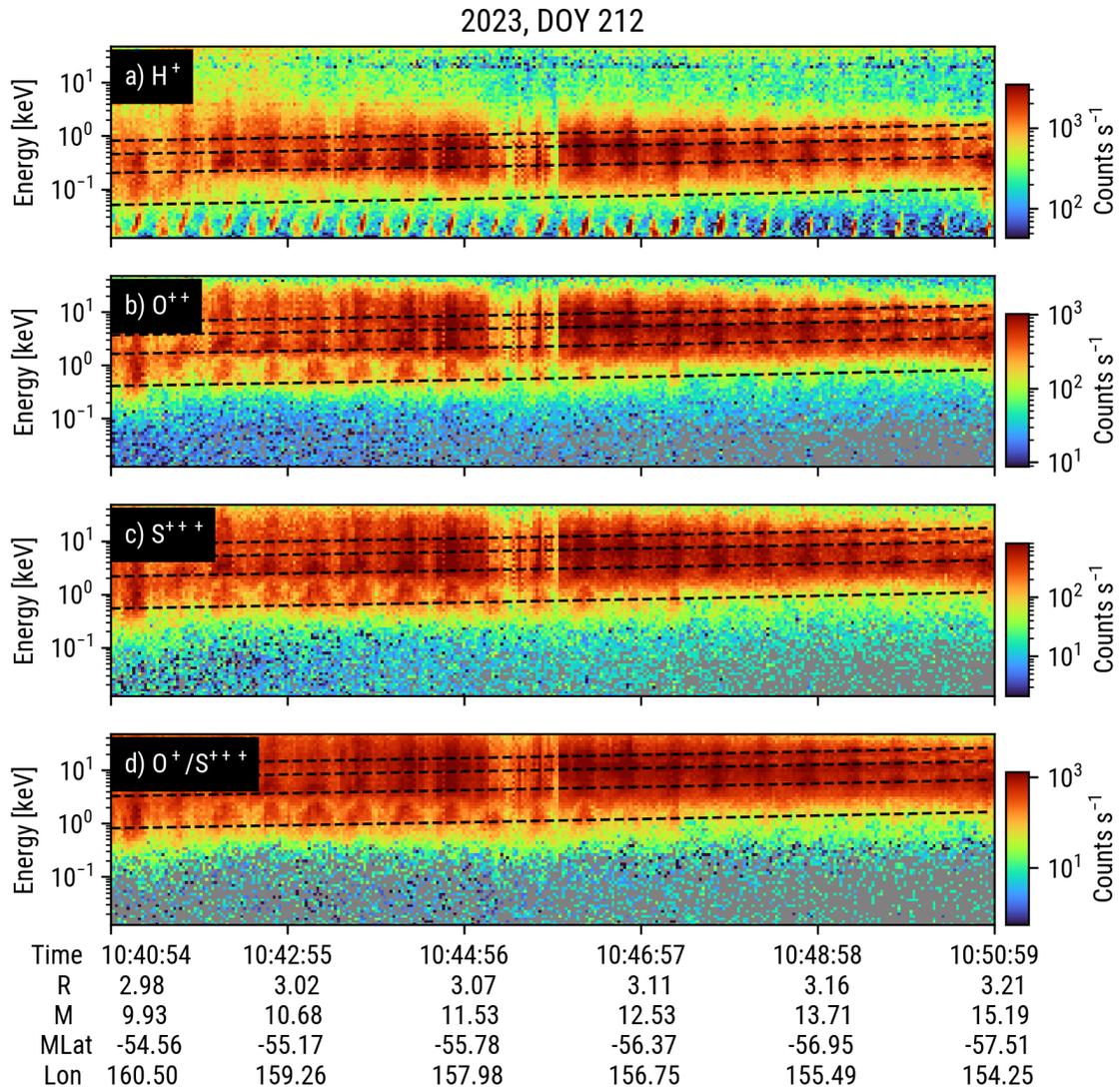

Figure S 28

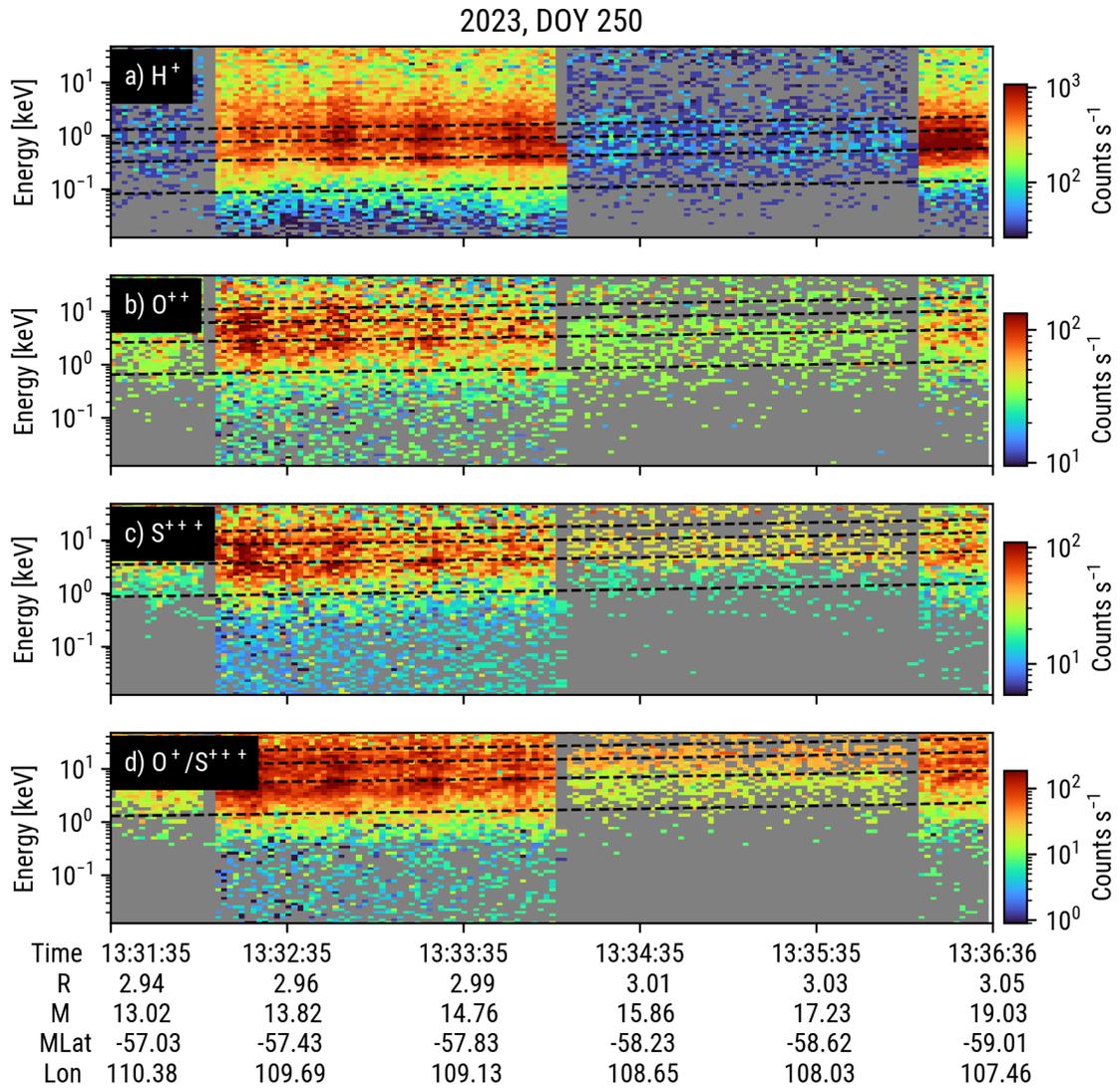

2023, DOY 250

Figure S 29

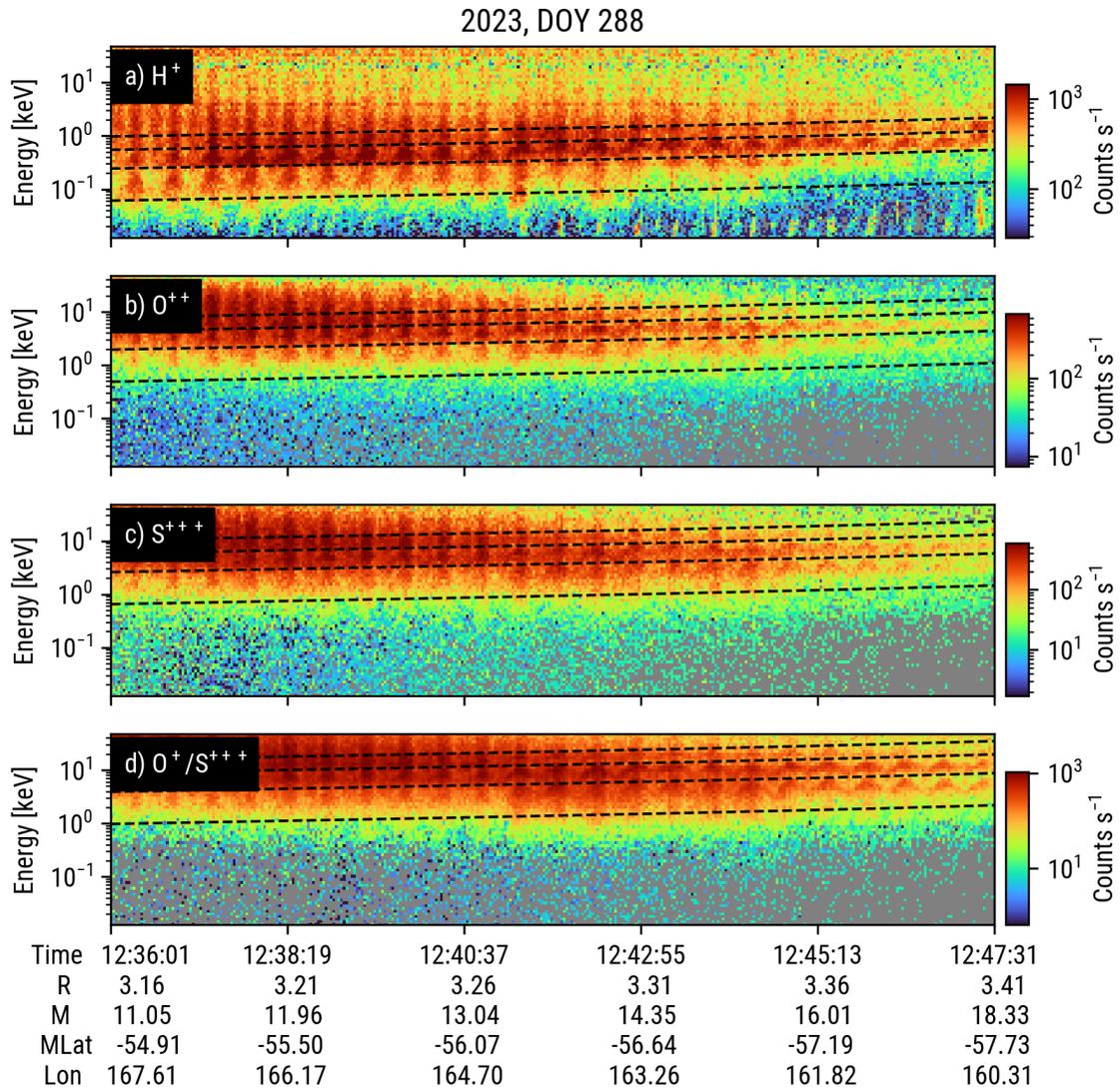

Figure S 30

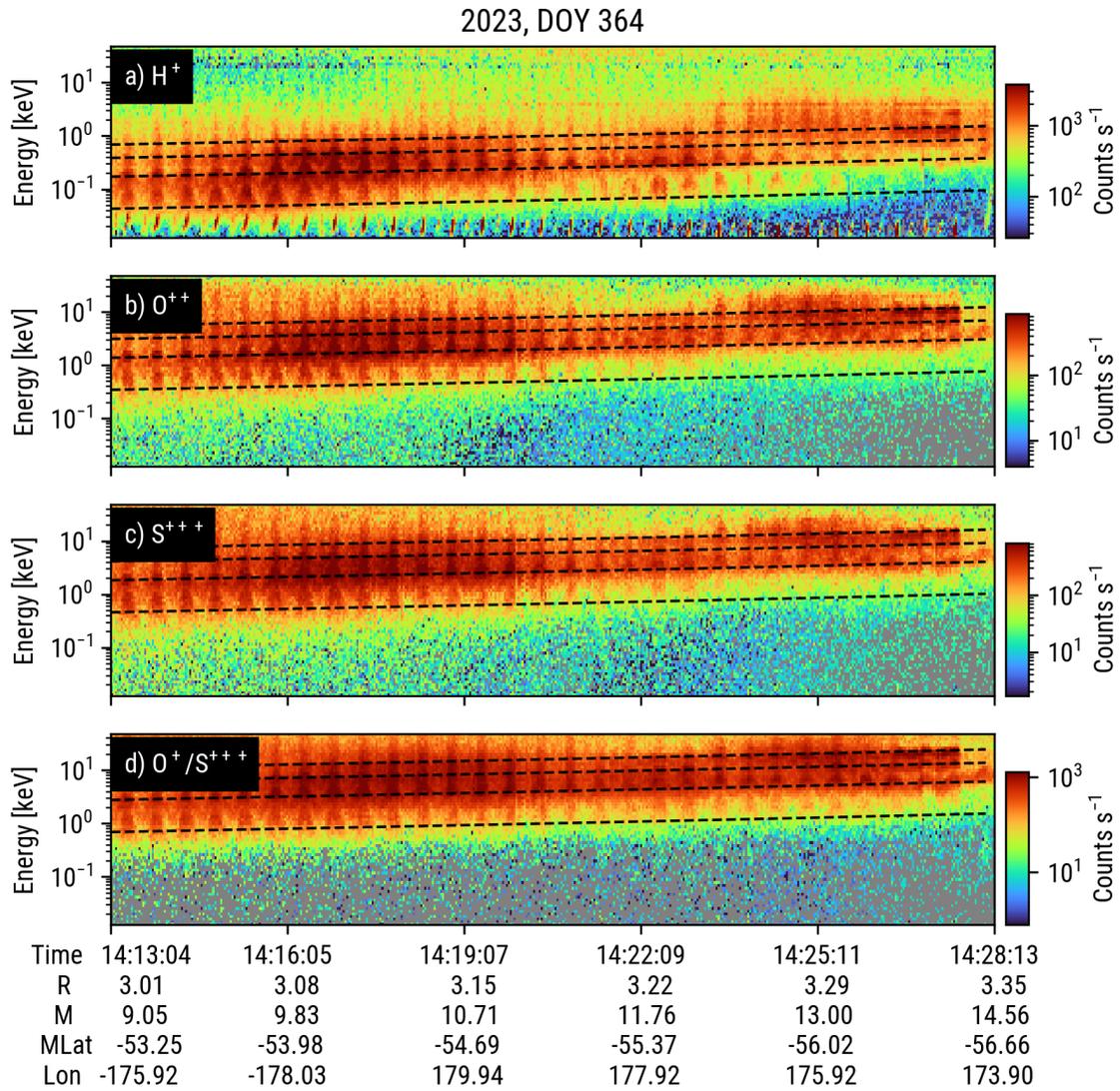

Figure S 31

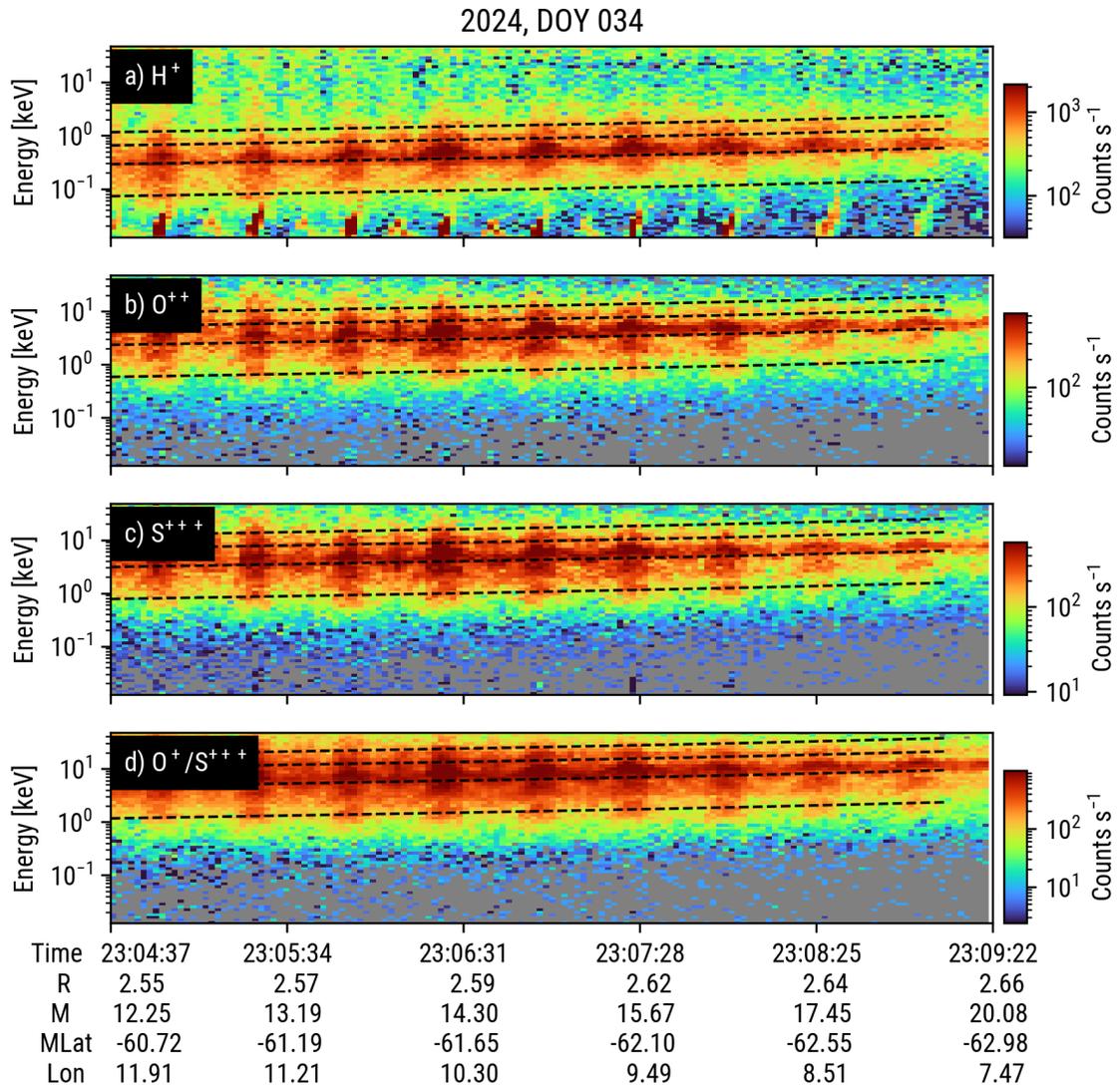

Figure S 32

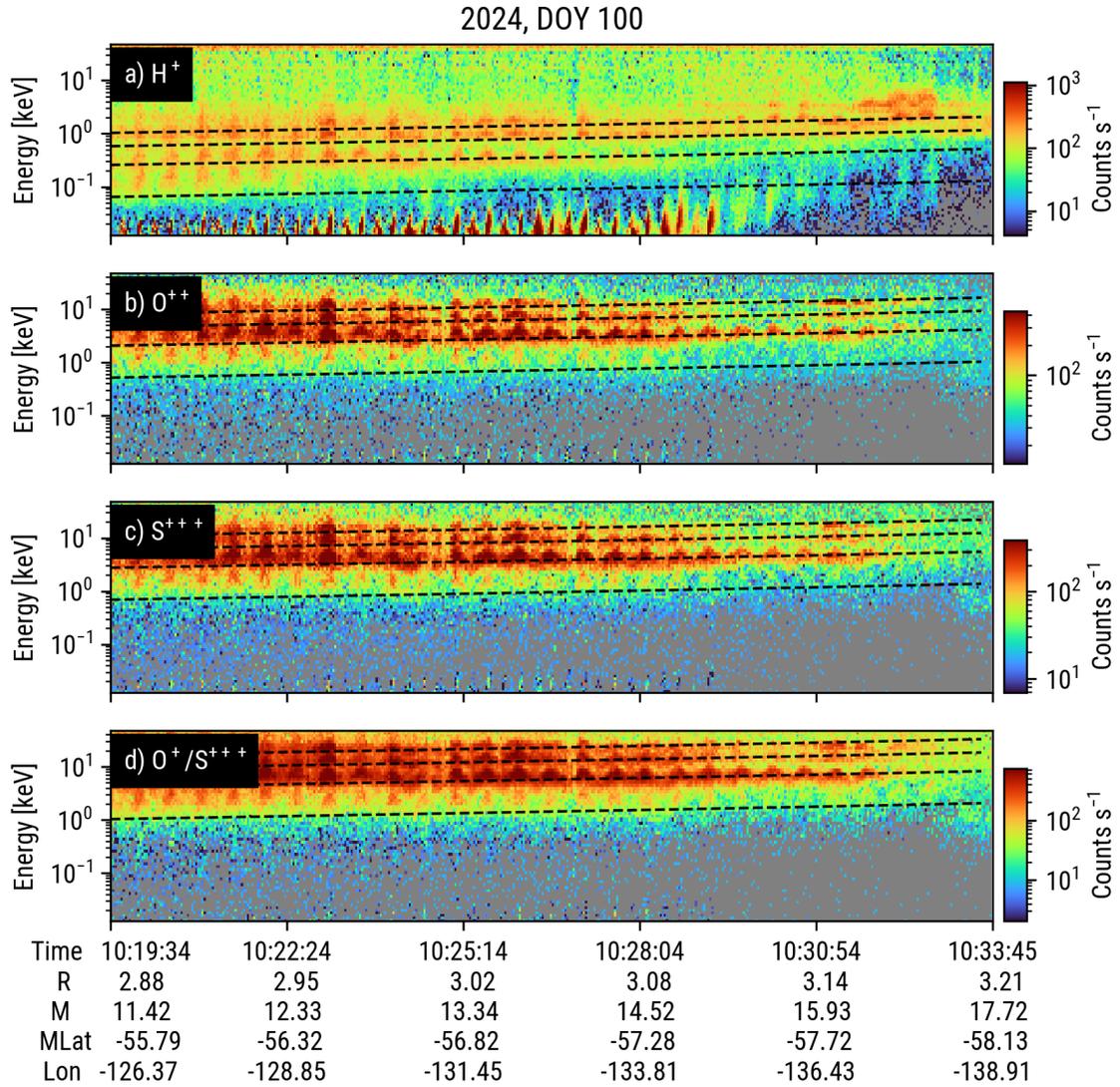

Figure S 33

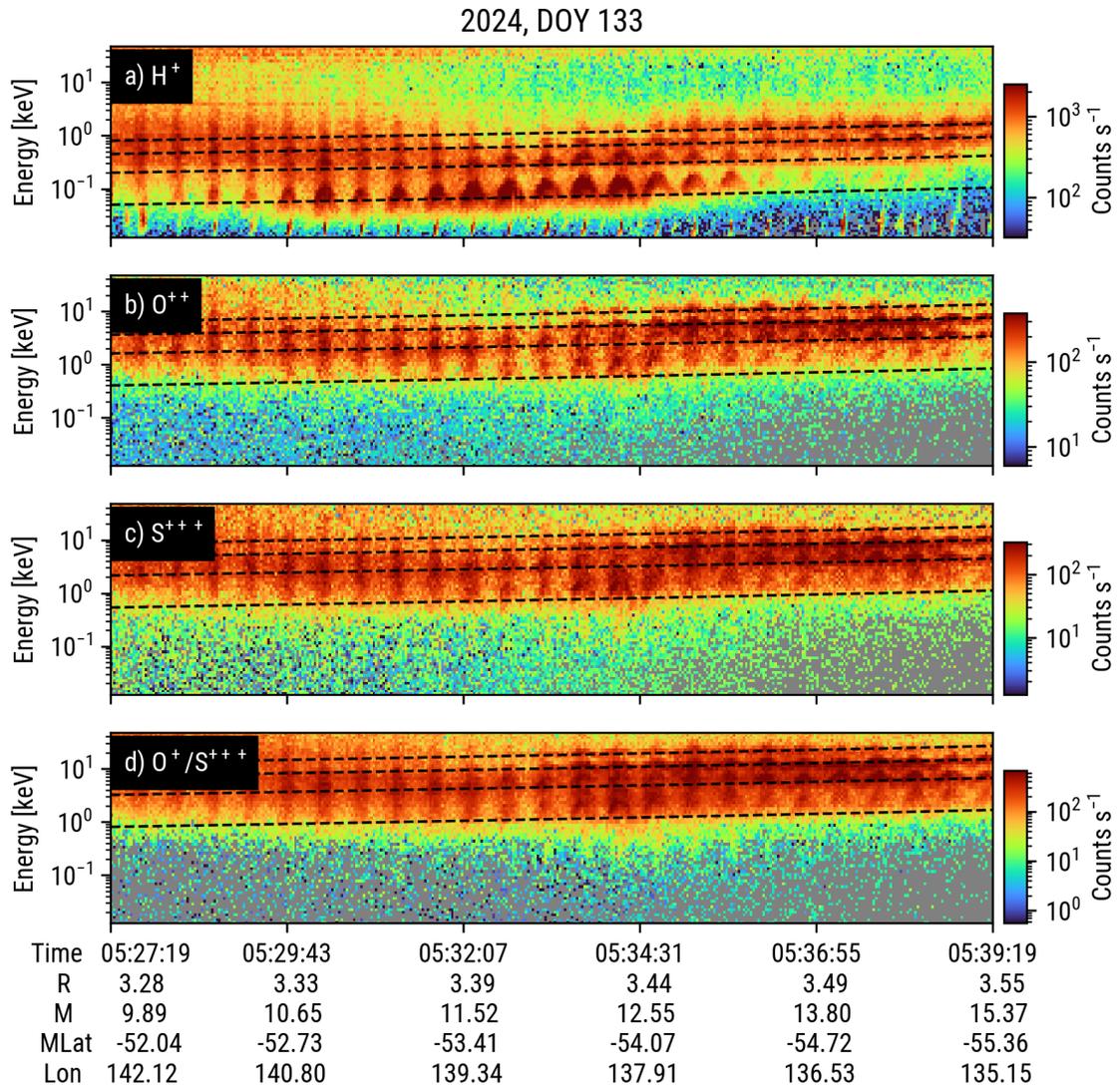

Figure S 34

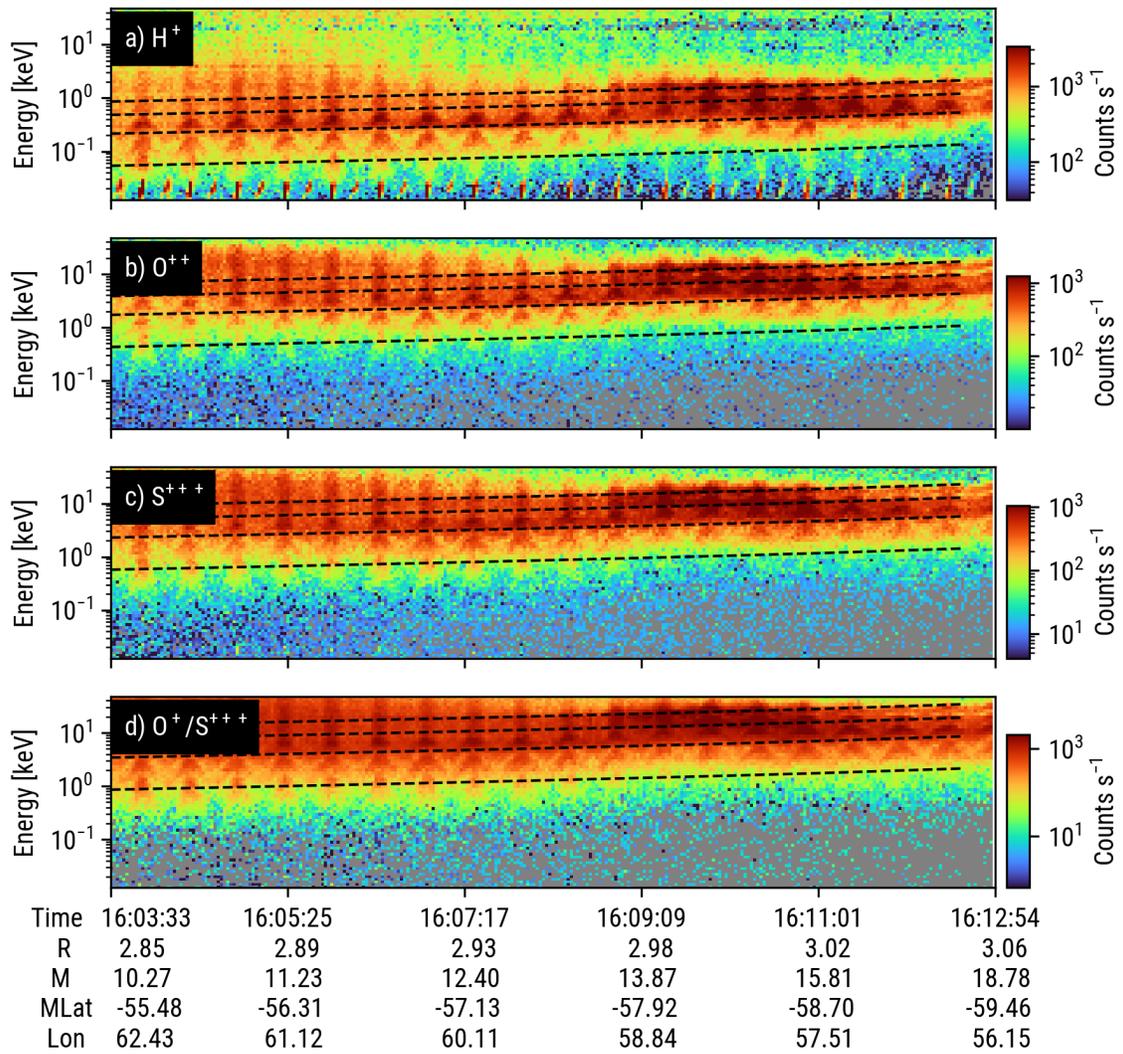

Figure S 35